\documentclass[fleqn,usenatbib]{mnras}

\usepackage[utf8]{inputenc}
\usepackage{graphicx}
\usepackage{hyperref}
\usepackage{mathtools}
\usepackage{comment}
\usepackage{amsmath}
\usepackage{ amssymb }
\usepackage{lipsum} 
\usepackage{float}
\usepackage{cleveref}
\usepackage[caption=false]{subfig}
\usepackage{placeins}
\usepackage{xcolor}
\usepackage{breqn}

\defcitealias{OBB}{OBB}
\defcitealias{KomatsuSeljak2001}{KS}

\def\thethreehundred{{\sc The Three Hundred}}
\def\GIZ{\textsc{Gizmo-SIMBA}}
\def\GM{\textsc{Gadget-MUSIC}}
\def\GX{\textsc{Gadget-X}}
\def\ahf{\textsc{AHF}}

\title[The Concentration-SZ Relation]{The impact of halo concentration on the Sunyaev Zel'dovich effect signal from massive galaxy clusters}

\author[Baxter et al.]{Eric J. Baxter,$^{1}$
Shivam Pandey,$^{2}$
Susmita Adhikari,$^{3}$
Weiguang Cui,$^{4}$
\newauthor
Tae-hyeon Shin,$^{5}$
Qingyang Li,$^{6}$
Elena Rasia$^{7,8}$
\\$^{1}$Institute for Astronomy, University of Hawai`i, 2680 Woodlawn Drive, Honolulu, HI 96822, USA
\\$^{2}$Department of Physics, Columbia University, New York, NY  10027, USA
\\$^{3}$Department of Physics, Indian Institute of Science Education and Research Pune, Maharashtra, 411045, India
\\$^{4}$Institute for Astronomy, University of Edinburgh, Royal Observatory, Edinburgh EH9 3HJ, United Kingdom
\\$^{5}$Department of Physics and Astronomy, Stony Brook University, Stony Brook, NY, 11794, USA
\\$^{6}$Department of Astronomy, School of Physics and Astronomy, Shanghai Jiao Tong University, Shanghai 200240, China
\\$^{7}$INAF - Osservatorio Astronomico di Trieste, via Tiepolo 11, I-34143 Trieste, Italy
\\$^{8}$Institute of Fundamental Physics (IFPU), Via Beirut, 2, 34151 Trieste TS, Italy
}

\begin{document}

\label{firstpage}

\pagerange{\pageref{firstpage}--\pageref{lastpage}}
\maketitle

\begin{abstract}
The Sunyaev Zel'dovich (SZ) effect is sensitive to the pressure of ionized gas inside galaxy clusters,  which is in turn  controlled largely by the gravitational potential of the cluster. Changing the concentration parameter describing the cluster mass distribution impacts the gravitational potential and thus 
the cluster SZ signal, with implications for cosmological and other analyses of SZ-selected clusters.  We investigate the concentration-SZ relation in theory and simulations.  We find that the impact of concentration on the inner SZ profile ($R \lesssim 0.75 R_{200c}$) can be captured with standard polytropic gas models.  However, we find that such models do a poor job of reproducing the outer SZ profiles ($R \gtrsim 0.75 R_{200c}$)  and the relation between the integrated SZ signal, $Y$, and concentration.  This disagreement results from a sharp truncation of the gas pressure profile near the splashback radius,  likely caused by virial shocks.  We develop a simple description of the truncation that leads to a good match with the simulated SZ profiles out to several $R_{200c}$ for clusters of varying mass and concentration, and that also accurately predicts the concentration-$Y$ relationship.  Finally, we determine how inference of the linear bias parameter and splashback radius for SZ-selected clusters can be biased by ignoring the concentration dependence of the SZ signal, finding that bias to the  former is essentially negligible, while bias to the latter can be as much as 2\%.
\end{abstract}  


\section{Introduction}

The Sunyaev Zel'dovich (SZ) effect arises from cosmic microwave background (CMB) photons inverse Compton scattering with hot electrons \citep{Sunyaev:1972}.  This process leads to a spectral distortion of the CMB that is sensitive to the line-of-sight integral of the electron gas pressure.   The SZ effect can be used to detect galaxy clusters because these objects contain large amounts of hot, ionized gas.  Cluster selection via SZ signal is especially powerful because it is effectively redshift-independent (unlike, e.g., optical and X-ray selection), and because the SZ signal correlates tightly with cluster mass.  CMB surveys like the South Pole Telescope (SPT), the Atacama Cosmology Telescope (ACT), and {\it Planck} have amassed large samples of galaxy clusters selected via their SZ signals \citep{Bleem:2015,Hilton:2021,Planck:clusters}.  Future CMB surveys like Simons Observatory \citep{SimonsObservatory} and CMB Stage 4 \citep{CMBS4} will detect orders of magnitude more clusters, enabling high precision cosmological constraints and studies of cluster physics.

Given the precision of current and future cosmological studies with SZ-selected clusters, it is important to build a comprehensive understanding of the connection between SZ signal and cluster mass profiles.  Total mass is the most important variable in determining this connection. The gas is roughly in hydrostatic equilibrium in the potential well of the cluster, leading to a tight relationship between the cluster mass and the gas pressure, which determines the SZ signal.  If the cluster mass distribution is self-similar and gas physics does not introduce any additional scales (e.g. the gas is adiabatic), then the gas properties will also be self-similar \citep{Kaiser:1986, Voit:2005}.   In this case, the integrated SZ signal from the cluster, $Y$, should scale as $M^{5/3}$, where $M$ is the cluster mass.   Additional physics, such as cooling \citep{Nagai:2006}, feedback from active galactic nuclei \citep{Gitti:2012}, non-thermal pressure support from bulk gas motions and turbulence \citep{Lau:2009,Shi:2015,Vazza:2018}, relativistic gas \citep{OBB}, and cosmic rays \citep{Pfrommer:2008,Ackermann:2014} complicate this simple picture, but the self-similar scaling still provides a reasonable match to both real and simulated clusters \citep[e.g][]{Voit:2005, Nagai:2006}.

Even at fixed total mass, though, there can be significant variation in the cluster gravitational potential, and thus variation in the equilibrium gas pressure and SZ signal.  One important form of such variation is concentration, defined as $c_{\Delta} = R_{\Delta}/R_{-2}$, where $R_{-2}$ is the radius at which the logarithmic slope of the cluster density profile reaches -2, and $R_{\Delta}$ is the spherical overdensity radius corresponding to an overdensity of $\Delta$.\footnote{The spherical overdensity radius, $R_{\Delta}$, of a halo at redshift $z$ is defined such that the average enclosed density within a sphere of radius $R_{\Delta}$ is equal to $\Delta$ times a reference density, $\rho(z)$: $M_{\Delta} = (4/3)\pi \Delta R_{\Delta}^3 \rho(z)$, where the spherical overdensity mass, $M_{\Delta}$, is the mass within $R_{\Delta}$, and $\rho(z)$ is often to chosen to be either the critical density or mean density of the Universe.  We will denote these two cases with subscripts `c' or `m,' respectively.  We use the virial overdensity (subscript `vir')  definition from \citet{Bryan:1998}.}
 On average, cold dark matter (CDM) halos are known to be well-described by the Navarro-Frenk-White (NFW) form \citep{NFW} with a concentration parameter that varies slowly with halo mass \citep[e.g.][]{Duffy:2008, Dutton:2014}; individual halos exhibit significant scatter around this average relation.  Since changes in  concentration impact the cluster potential, one would expect these changes to also impact cluster SZ signals \citep{KomatsuSeljak2001}.  The goal of this work is to investigate the relationship between concentration and SZ signal in theory and simulations, and to explore the implications of this relationship for analyses of SZ-selected clusters.  

Because $Y$ is closely related to the observables used to select clusters, it is particularly important to understand the $Y$-$c_{\Delta}$ relation.  The integrated SZ signal within commonly chosen apertures (e.g. $R_{200c}$ and $R_{500c}$) receives  contributions from the cluster outskirts, where several physical effects can become important.  For one, at large halo-centric distances, the contributions from nearby halos or halos projected along the line of sight can be significant (i.e. the two-halo term in the language of the halo model, \citealt{Vikram:2017,Hill:2018}).   Another phenomenon modifying the outer SZ profiles of halos is the presence of large-scale shocks.  Gas accreting onto clusters typically experiences one or more shocks on its way to becoming part of the intracluster medium.  Such shocks are predicted by self-similar collapse models \citep[e.g.][]{Bertschinger:1985, Shi:2016} to coincide roughly with the cluster splashback radius \citep{Diemer:2014}.  More recently, hydrodynamical simulations have shown that clusters can be surrounded by so-called merger-accelerated accretion shocks that are initiated by halo mergers, and driven to large cluster-centric radius \citep{Zhang:2020}.  Because shocks change the gas pressure, they can leave imprints on the SZ profiles of galaxy clusters \citep{Molnar:2009,Baxter:2021}.  We will show below that both shocks and the two-halo term have significant impact on the $Y$-$c_{\Delta}$ relation.

Correlation between $Y$ and concentration will have implications for cosmological and other analyses of SZ-selected galaxy clusters.  Assuming one selects clusters on $Y$ (or some closely related quantity, like SZ signal-to-noise), the $Y$-$c_{\Delta}$ relation will modify the  concentration distribution of the resultant cluster samples.  In this case, inferred quantities which also depend on concentration may be biased by ignoring the $Y$-$c_{\Delta}$ relation.  We will consider two such quantities here: the linear bias parameter characterizing the clustering of clusters, and the splashback radius.  In the case of the former, the linear bias depends on concentration as a result of halo assembly bias \citep{Wechsler:2006,Dalal:2008}.  Thus, analyses which attempt to infer cluster masses from the amplitude of clustering \citep[e.g.][]{Holder:2006, Hu:2006,Baxter:2016,Chiu:2020} using SZ-selected cluster samples could be biased if the $Y$-$c_{\Delta}$ connection is ignored \citep{Wu:2008}.  On the other hand, the splashback radius corresponds to the first apocenter of accreting dark matter \citep{Diemer:2014}.  This radius is anti-correlated with accretion rate, which is in turn known to anti-correlate with halo concentration.  Thus, as seen in \citet{Diemer:2014}, the splashback radius has a positive correlation with halo concentration.  Consequently, measurements of the splashback radius with SZ-selected cluster samples \citep[e.g.][]{Shin:2019} could also be biased by correlation between $Y$ and $c_{\Delta}$.  Understanding the amplitude of this effect is particularly important given the apparent discrepancy between some measurements of the splashback radius and theoretical predictions \citep[e.g.][]{More:2016,Baxter:2017}, although the source of these discrepancies likely lies in issues with optical cluster selection \citep{Busch:2017}.

Several previous works have considered the connection between SZ signal and concentration. \citet{KomatsuSeljak2001} developed a model for cluster pressure profiles that explicitly includes dependence on $c_{\Delta}$, showing that this dependence could lead to a departure from the self-similar prediction for the $Y$-$M$ relation; we will consider this model in detail below.  \citet{Wu:2008} considered how correlation between cluster SZ signals and concentration could impact cluster mass calibration via clustering.  \citet{Green:2020} explored the contributions of accretion history and concentration to scatter in the $Y$-$M$ relation.  \citet{Wadekar:2022} investigated whether the halo concentration could be used to reduce scatter in the $Y$-$M$ relation. \citet{Lee:2022} explored how halo concentration impacts SZ signals in the Illustris simulations, using a simple fitting function to measure this relation. Our work provides physical insight into the origin of the correlation seen in \citet{Lee:2022}.

The plan of the paper is as follows.  In \S\ref{sec:simulations} we describe The Three Hundred project hydrodynamical zoom-in simulations of massive galaxy clusters \citep{The300}.  In \S\ref{sec:models} we develop several models for describing how the cluster SZ signal changes with concentration.  \S\ref{sec:resultsI} presents our investigation into the impact of concentration on the inner SZ profile.  In \S\ref{sec:resultsII}, we consider the outer SZ profiles and the relation between integrated SZ signal and concentration.  One of our main results is to show that truncation of the halo pressure profile at a radius different from $R_{200c}$ is needed to capture the $Y$-$c_{\Delta}$ relation.   In \S\ref{sec:resultsIII}, we fit our model for the $Y$-$M$-$c_\Delta$ relation to simulations, and use the results of these fits to assess how measurements of the linear bias parameter and splashback radius with SZ-selected clusters can be biased.  We conclude in \S\ref{sec:discussion}.

\begin{figure}
    \centering
    \includegraphics[scale=0.4]{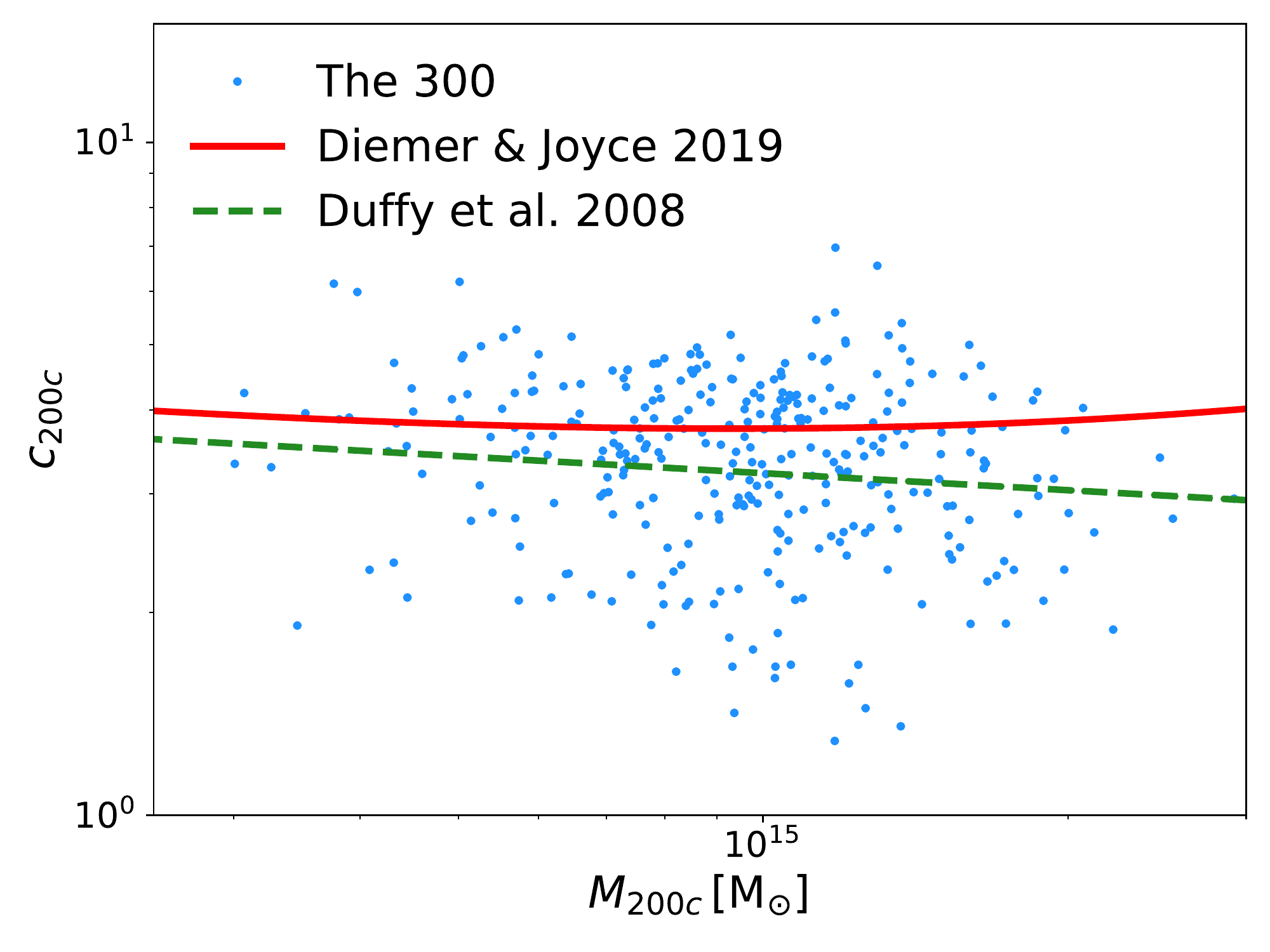}    
    \caption{The mass-concentration distribution of the clusters in The300 simulations (blue points).  Curves show two fitting functions for the mass-concentration relation.}
    \label{fig:mass-conc}
\end{figure}

\section{Simulation data products}
\label{sec:simulations}

The Three Hundred Project\footnote{\url{https://the300-project.org}}~\citep[][hereafter The300]{The300} performed hydrodynamical re-simulations of the most massive 324 galaxy clusters at $z = 0$ from the MultiDark Planck 2~\citep[MDPL2,][]{Klypin2016}\footnote{\url{https://www.cosmosim.org/cms/simulations/mdpl2}} $N$-body simulation. Each cluster has its own initial conditions which are re-generated to reduce the computation cost: the high-resolution region which includes dark matter (DM) and gas-particle masses with masses of $m_{\rm DM}\simeq 12.7 \times 10^8\ h^{-1}\mathrm{M_{\odot}}$ and $m_{\rm gas} \simeq 2.36\ \times 10^8\ h^{-1}\mathrm{M_{\odot}}$ respectively, has a radius of 15 $h^{-1}\mathrm{Mpc}$ centred around the selected galaxy cluster at $z = 0$.  Outside of the high-resolution region, the simulation box is filled with multi-level low-resolution DM particles. The same {\it Planck} 2016 cosmological parameters  \citep{Planck2016} as the parent MDPL2 simulation are adopted. This set of initial conditions was run with three different flavors of baryon model: \GM\ \citep{Sembolini2013}, \GX\ \citep{Rasia2015}, and \GIZ\ \citep{Dave2019, Cui2022}. We only use the \GX\ results for this work, which will be briefly detailed in the following subsection. Benefiting from the large number of high-mass clusters, different physical models and very large high-resolution region, these simulated clusters have been used for several studies related to this work: cluster profiles \citep{Mostoghiu2019,Li2020,Baxter:2021,Sayers:2023}, back-splash galaxies \citep{Arthur2019,Haggar2020,Knebe2020}, mass-concentration relation \citep{Darragh-Ford:2023}, and cluster dynamical state \citep{DeLuca2021,Capalbo2021,Zhang2022,Li2022}.

\subsection{The \GX\ Baryon model}

\GX\ is developed on top of the gravity solver of the {\sc GADGET3} Tree-PM code (an updated version of the {\sc GADGET2} code; \citealt{Springel2005}). The improvements include  artificial thermal diffusion, time-dependent artificial viscosity, a high-order Wendland C4 interpolating kernel and a weak-up scheme \citep{Beck2016} for the smoothed particle hydrodynamics; stellar evolution with consideration of mass-dependent lifetimes of stars \citep{Padovani1993} implemented in \citet{Tornatore2007}; the production and evolution of 15 different elements coming from SNIa, SNII and AGB stars with metallicity-dependent radiative cooling \citep{Wiersma2009}; an improved black hole (BH) growth and implementation of active galactic nuclei (AGN) feedback \citep{Steinborn2015}. The BH accretion is based on the Eddington-limited Bondi accretion with individual accretion rates for hot (boost factor $\alpha = 10$) and cold ($\alpha = 100$) gas respectively, while the active galactic nuclei (AGN) feedback, modelling both the mechanical and radiative modes, is implemented as thermal feedback. We refer readers to \cite{The300,Cui2022} for detailed descriptions of the model differences; \cite{Zhang2022} for the dynamical state differences; Li et al. 2023 (in prep.) for the baryon profile comparisons between these different runs. 

\subsection{Data selection}

Following our previous work on signatures of shocks in cluster SZ profiles \citep{Baxter:2021}, all the central clusters at $z = 0.193$ are selected for analysis in this work.  This restriction in redshift can be viewed as a limitation of our analysis, but it does not impact our main results.  Moreover, many of the very high-mass clusters found by ground-based CMB surveys have similar redshifts \citep[e.g.][]{Bleem:2015}.  
The halos in the high-resolution regions are initially identified by the Amiga Halo Finder~\citep[\ahf,][]{Knollmann2009} with an overdensity of $200\rho_{\rm crit}$, where $ \rho_{\rm crit}$ is the critical density of the Universe at the corresponding redshift. The values of $M_{500c}$ and $R_{500c}$ for each halo are calculated from the halo density profiles.   A typical mass of a cluster in The300 sample is $M_{200c} \approx 10^{15}\,\mathrm{M_{\odot}}$.  

\subsection{Concentration measurements}

To determine the concentration of each cluster, we fit an NFW profile \citep{NFW} to the total density:
\begin{equation} \label{eq:NFW}
    \rho_{\rm NFW}(r) = \frac{\rho_0}{\frac{r}{r_s}(1+\frac{r}{r_s})^2},
\end{equation}
where $\rho_0$ sets the normalization of the density profile and $r_s$ is the scale radius of the halo.\footnote{We will typically use lowercase $r$ to refer to the 3D distance to the cluster center, and uppercase $R$ to refer to the projected distance to the cluster center from the line of sight.} Both are free parameters in the fitting. The concentration is given by $c_{\Delta} = R_{\Delta}/r_s$. Note that the total central density of a simulated halo with baryons tends to be cuspy, which can not be fit by an NFW profile \citep[see][for example]{Cui2014, Schaller2015}. Therefore, we excluded the inner $0.025R_{200c}$ in radius when fitting.

Fig.~\ref{fig:mass-conc} shows the relation between halo mass and concentration for the 324 clusters in our sample.   For comparison, we show the predictions of two common fitting functions for the mass-concentration relation: \citet{Duffy:2008} and \citet{Diemer:2019}.  Note that these fitting functions were obtained using dark matter-only simulations.  In the absence of strong feedback, baryonic cooling results in enhanced baryon density in halo cores leading to increased concentration; however, strong feedback from AGN and supernovae removes gas from the halo cores, leading to a \textit{decrease} in concentration \citep{Duffy:2010}.  In general, though, the impact of baryonic physics on the concentrations of cluster-scale halos  is relatively small, typically less than ten percent \citep{Duffy:2010}.

\begin{figure}
    \centering
    \includegraphics[scale=0.45]{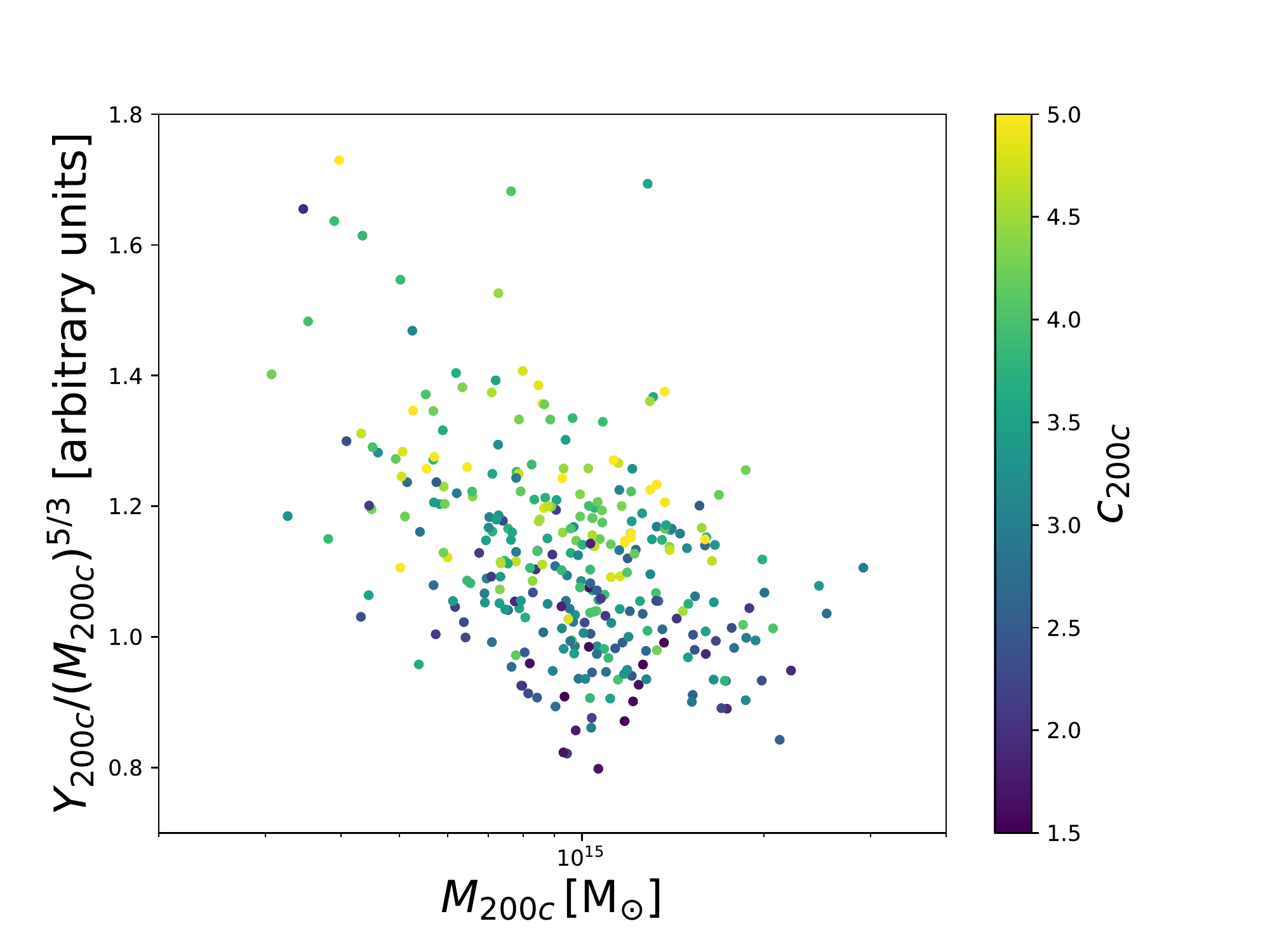}
    \caption{The relationship between $Y_{200c}$, $M_{200c}$, and $c_{200c}$ in The300 simulations at $z = 0.193$.  Points correspond to individual clusters from the simulation.  We scale the $y$-axis by the self-similar expectation, $Y_{200c} \propto M_{200c}^{5/3}$.  There is a clear trend for clusters with higher concentration to have higher $Y_{200c}$ at fixed halo mass.}
    \label{fig:YM_scatter}
\end{figure}

\subsection{SZ profile measurements}

The SZ maps used in this work are described in \cite{Baxter:2021}.  The maps cover a 12 ${\rm Mpc}/h$ radius region around the center of the high-resolution region at $z=0.1913$.  We use the {\sc PyMSZ} package\footnote{\url{https://github.com/weiguangcui/pymsz}, see \citet{The300} for more detailed descriptions.} to generate SZ maps at resolution corresponding roughly to $2.5\arcsec$.  

We measure the radial SZ profiles of all 324 simulated clusters in 50 bins of projected radius from 0.05 to 12 Mpc.  The profile measurement in each radial bin is the average of all SZ map values within that bin.  We also measure the 3D pressure profiles from the simulations using 150 radial bins from 0.01 Mpc to 10 Mpc.   Integrating the measured SZ profiles out to $R_{\Delta}$ yields the integrated SZ signal, $Y_{\Delta}$ (see Eq.~\ref{eq:integratedY}).

Fig.~\ref{fig:YM_scatter} shows the relationship between $Y_{200c}$ and $M_{200c}$ for individual halos, colored by $c_{200c}$.  We have scaled the $y$-axis in the plot by the self-similar expectation, $Y \propto M^{5/3}$.  It is clear from the figure that there is some correlation between $Y_{200c}$ and $c_{200c}$ at fixed $M_{200c}$, with clusters with larger concentration tending to have higher $Y_{200c}$.  Understanding this correlation is one of the main goals of this paper.  The scaling of $Y_{200c}$ with $M_{200c}$ departs slightly from the self-similar scaling (i.e. it is flatter than $Y \propto M^{5/3}$), in agreement with e.g. \citet{Nagai:2006}.

\section{Modeling the impact of concentration on SZ signal}
\label{sec:models}

We consider several approaches to modeling the response of the SZ signal to changes in the concentration.  Our main focus is polytropic gas models (\S\ref{sec:polytrope}), which we supplement with a simple prescription for non-thermal pressure support and the contributions from nearby halos (i.e. the two-halo term).  In \S\ref{sec:resultsII} we will further modify the polytropic model to improve agreement with the simulations.   We also consider a simple model for the $Y_{\Delta}$-$c_{\Delta}$ relation based on total gravitational potential energy (\S\ref{sec:potential_argument}), as well as an approach where the relation between SZ profile and concentration is described via flexible fitting functions (\S\ref{sec:fitting_functions}). 

\subsection{Total potential argument}
\label{sec:potential_argument}

$Y_{\Delta}$ receives contributions from the electron pressure throughout the cylindrical volume of radius $R_{\Delta}$ oriented along the line of sight to the cluster.  Since the outer pressure profile of the cluster declines steeply with radius, we can reasonably approximate this cylinder with a sphere of radius $R_{\Delta}$ centered on the cluster.  In this case, $Y_{\Delta}$ is proportional to the total thermal energy of the gas within $R_{\Delta}$.  

One expects the thermal energy of the gas to be roughly proportional to the total potential energy of the halo \citep{Voit:2005, Fujita:2018}, with constant of proportionality roughly $\Omega_b/\Omega_m$. 
The potential energy of the cluster can be computed assuming that the density profile is given by the NFW form, yielding a  relation between $Y_{\Delta}$ and $c_{\Delta}$:
\begin{multline}
  Y_{\Delta} \propto M^{5/3} c_{\Delta} \left[ \ln(1+c_{\Delta}) - c_{\Delta}/(1+c_{\Delta})\right]^{-2}  \\
  \left[(1/2) - \frac{2(1+c_{\Delta})\ln(1+c_{\Delta})+1}{2(1+c_{\Delta})^2}\right]. 
\end{multline}
Note that the self-similar scaling ($Y \propto M^{5/3}$) emerges naturally from this argument.  We will use this simple model as a basis for comparison. 

\subsection{Fitting functions}
\label{sec:fitting_functions}

A common form used to fit the pressure profiles of clusters is the generalized NFW profile:
\begin{equation}
P(r) = P_0 \left( \frac{x}{x_c}\right)^{\gamma}\left[1 + \left(\frac{x}{x_c} \right)^{\alpha} \right]^{-\beta},
\end{equation}
where $x = r/R_{\Delta}$, and $P_0$, $\alpha$, $\beta$, $\gamma$, and $x_c$ are fitting parameters.  \cite{Battaglia:2012} fit cluster profiles from hydrodynamical simulations to this form, fixing $\alpha = -1.0$ and $\gamma = -0.3$, while allowing $P_0$, $x_c$ and $\beta$ to vary with mass and redshift according to
\begin{equation}
\label{eq:Mz_dependence}
    A(M_{200c},z) = A_0 \left( \frac{M_{200c}}{10^{14}\,\mathrm{M_{\odot}}} \right)^{\alpha_m} (1+z)^{\alpha_z},
\end{equation}
where $A$ generically represents a parameter from the set $\{P_0, x_c, \beta \}$.

\citet{Lee:2022} introduce additional freedom into Eq.~\ref{eq:Mz_dependence},  allowing for (1) a break in the power law scaling with mass (i.e. different values of $\alpha_m$ at high and low halo mass), and (2) dependence of the parameters on halo concentration. Introduction of a break in the power law is motivated by the impact of baryonic feedback on the $Y$-$M$ relation, and is expected to be important for $M_{200c} \lesssim 10^{14}\,\mathrm{M_{\odot}}$.  Since we are interested here in the SZ-concentration relation at high halo masses $M_{200c} > {\rm few} \times 10^{14}\,\mathrm{M_{\odot}}$, we do not consider broken power laws in mass.  We do, however, consider additional dependence of the parameters on halo concentration following the \citet{Lee:2022} form:
\begin{equation}
A(M,z,c) = A_0 \left( \frac{M}{10^{14}\,\mathrm{M_{\odot}}} \right)^{\alpha_m} (1+z)^{\alpha_z} (c/10)^{\alpha_c}.
\end{equation}
This form allows the amplitude of the pressure profile (via $P_0$) as well as its shape (via $x_c$ and $\beta$), to vary with halo concentration.

\subsection{Polytropic gas model}
\label{sec:polytrope}

The main focus of our analysis is polytropic gas models, relying heavily on the treatment in  \citet{KomatsuSeljak2001} (hereafter \citetalias{KomatsuSeljak2001}).   We describe their model here briefly, but refer readers to \citetalias{KomatsuSeljak2001} for more details.  We will use the polytropic model to derive a prediction for the SZ profile of a halo as a function of its mass and concentration (and redshift).  For a polytrope, the gas pressure at radius $r$, $P_{\rm gas}(r)$, is related to the gas density, $\rho_{\rm gas}(r)$, via
\begin{equation}
P_{\rm gas}(r) \propto \rho_{\rm gas}(r) T_{\rm gas}(r) \propto \rho_{\rm gas}(r)^{\gamma},
\end{equation}
where $T_{\rm gas}(r)$ is the temperature  of the gas and $\gamma$ is the polytropic index. 

The gas is assumed to reside in a total mass distribution described by an NFW profile (Eq.~\ref{eq:NFW}), which is a function of the scale radius $r_s$.  Note that the NFW profile can be expressed in terms of the halo mass and concentration (our main parameters of interest) using $r_s = R_{\Delta}/c_{\Delta}$, where $R_{\Delta}$ is related to $M_{\Delta}$ and the cluster redshift as described previously, and 
\begin{equation}
    \rho_0 = \frac{M_{\Delta}}{4\pi R_{s}^3 \left( \ln (1+c) - c/(1+c)\right)}.
\end{equation}
This expression ensures that the volume integral of the density profile within $R_{\Delta}$ yields $M_{\Delta}$.

Assuming that the gas physics does not introduce a new scale into the problem, the gas density profile can be written in self-similar form as
\begin{equation}
    \rho_{\rm gas} (r) = \rho_{\rm gas}(0) u(r/r_s),
\end{equation}
where $u(r/r_s)$ is a function to be determined.  Combining this expression with the polytropic relation, we have
\begin{equation}
    P_{\rm gas}(r) = \frac{k_B T(0) \rho_{\rm gas}(0)}{\mu} u(r/r_s)^{\gamma},
\end{equation}
where $T(0)$ is the central gas temperature and $\mu$ is the mean mass of the gas particles.  The mean mass is given by $\mu = 4/(5X + 3) \approx 0.588$, where $X \approx 0.76$ is the primordial Hydrogen mass fraction.

Hydrostatic equilibrium then requires that 
\begin{equation}
   \gamma \frac{k_BT(0)}{u(r)} \frac{du}{dr} = -G \frac{M(r)}{r^2}, \label{eq:HSE}
\end{equation}
where $M(r)$ is the total mass enclosed within $r$ (given by the NFW profile).  Since $\rho_{\rm gas}(0)$ drops out of Eq.~\ref{eq:HSE}, the only free parameters needed to solve for $u(r)$ are $T(0)$ and $\gamma$ once $M_{\Delta}$, $c_{\Delta}$ and the cluster redshift have been specified.

\citetalias{KomatsuSeljak2001} fix $\gamma$ and $T(0)$ by requiring that (up to an overall scaling), the gas density profile traces the dark matter profile in the cluster outskirts, as motivated by the results of hydrodynamical simulations.  This amounts to the requirement that at some radius, $r_*$, in the halo outskirts, the logarithmic slopes of the gas and dark matter density profiles match:
\begin{equation}
    \frac{d \ln \rho}{dr} \bigg\rvert_{r = r_*} = \frac{d \ln \rho_{\rm gas}}{dr} \bigg\rvert_{r = r_*}.
\end{equation}
This equation fixes $T(0)$.  The polytropic index is then fixed by requiring the above equation to hold (at least approximately) over a range of $r_*$ from $0.5R_{\Delta}$ to $2R_{\Delta}$ (see \citetalias{KomatsuSeljak2001} for more details). 
 
The above arguments fix both $\gamma$ and $T(0)$, and thus the temperature profile of the gas.  However, there is additional freedom to choose the amplitude of the density profile, $\rho_{\rm gas}(0)$, since this term drops out of the hydrostatic equilibrium equation.  Note that $\rho_{\rm gas}(0)$ is necessary to fully specify the pressure profile $P_{\rm gas}(r)$, which is the relevant quantity for determining the SZ profile.  Here, we fix the normalization of the gas density and pressure profiles by requiring that the baryon fraction within $R_{200c}$ be equal to the cosmic mean, i.e. $\Omega_b/\Omega_m$ \citep{Komatsu:2002:MNRAS:}.  This is reasonable for cluster-scale halos \citep{Morandi:2015}.  We refer to the resultant model for the gas pressure as $P_{\rm KS}(r)$.

\begin{figure*}
    \centering
    \includegraphics[scale=0.5]{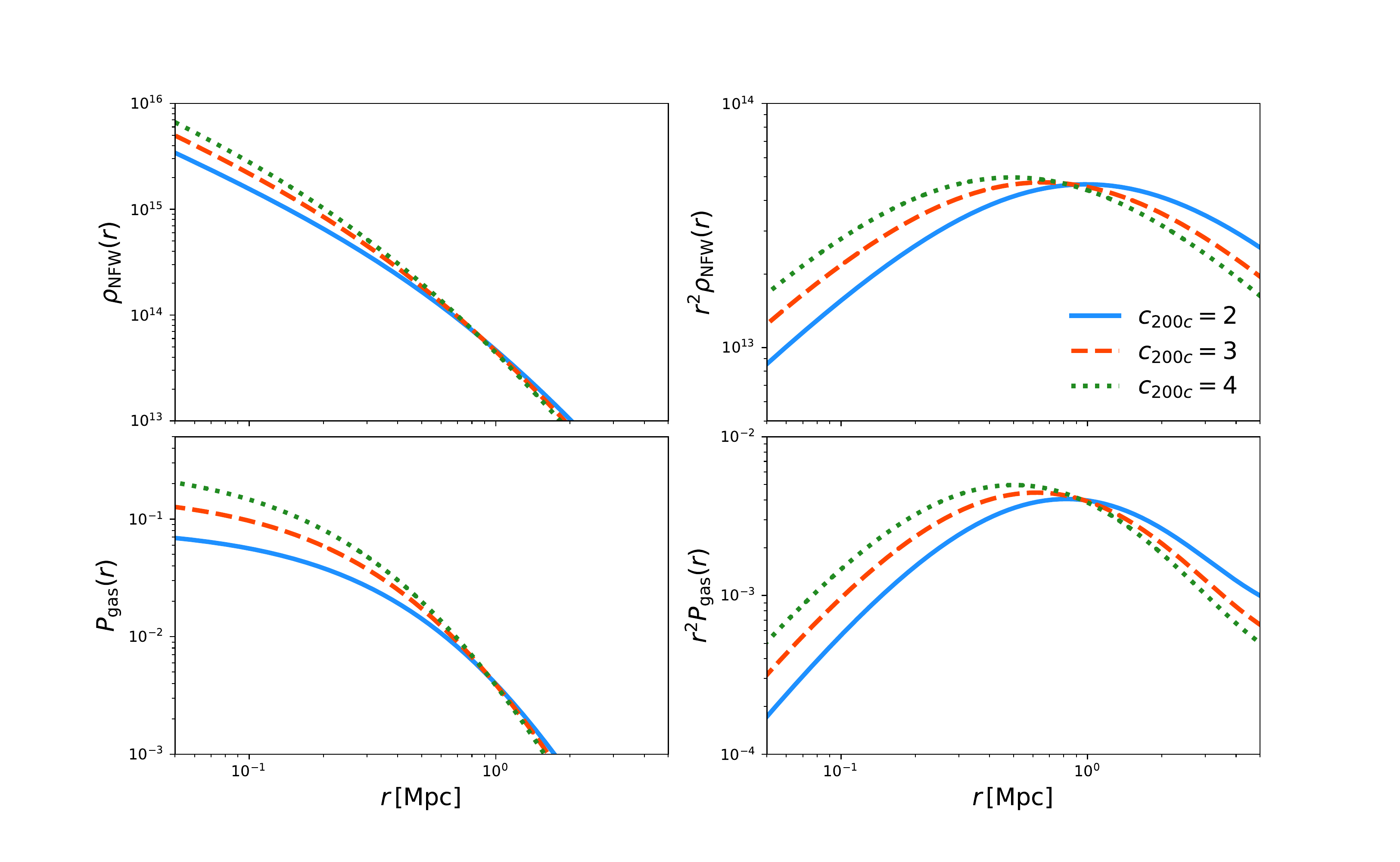}
    \caption{Variation in the cluster density profile, $\rho_{\rm NFW}(r)$ (top panels), and gas pressure profile, $P_{\rm gas}(r)$ (bottom panels), with concentration, as predicted by the \citetalias{KomatsuSeljak2001} model described in \S\ref{sec:polytrope}.  Profiles are plotted as a function of distance from the cluster center, $r$.  In the right column, we scale the pressure and density profiles by $r^2$ to make the variation in the outer profile clearer.}
    \label{fig:polytrope}
\end{figure*}

We show $P_{\rm KS}(r)$ as a function of $c_{\Delta}$ in Fig.~\ref{fig:polytrope}.  In this figure, we have fixed the halo mass to be $M_{200c}  = 10^{15}\,\mathrm{M_{\odot}}$, typical for the clusters in The300.  At fixed concentration, the inner pressure profiles (bottom panels) are always flatter than the dark matter density profiles (top panels).  This is because of the additional thermal pressure support provided by the gas.  As concentration is increased, the inner density profile increases in amplitude.  For $r \ll r_s$, the density profile scales like 
\begin{eqnarray}
\rho_{\rm NFW}(r \ll r_s) &\approx &\frac{\rho_0 r_s}{r} \nonumber \\
&=& \frac{R_{\Delta}}{r}\frac{\Delta \rho(z)}{3}\frac{c_{\Delta}^2}{\ln(1+c_{\Delta}) - c_{\Delta}/(1+c_{\Delta})} , \nonumber \\
\end{eqnarray}
which is an increasing function of concentration.  At the same time, the density profile at fixed $r < r_s$ becomes steeper because $r_s$ (the location where the density profile reaches a logarithmic slope of $-2$) decreases as $c_{\Delta}$ is increased.  The steeper and larger amplitude density profile means that the gas pressure profile must also become steeper and larger in amplitude to maintain hydrostatic equilibrium.  At fixed halo mass and halo baryon fraction, these increases in pressure in the inner profile are compensated by decreases in the outer profiles, as can be seen most clearly in the right column of Fig.~\ref{fig:polytrope}.

\subsubsection{Non-thermal pressure support and other effects}
We do not expect the KS model to exactly match real clusters or simulations because of the effects of e.g. radiative cooling, star formation and non-thermal pressure support from bulk and turbulent gas motion.  These effects can reduce the thermal pressure by tens of percent \citep{Nagai:2006,Shi:2015,Vazza:2018}.  

We account for non-thermal pressure support following \citet{Nelson:2014}.  That work used simulations to develop fitting functions for the nonthermal pressure component within galaxy clusters as a function of halocentric distance and accretion rate.  They found that when the cluster-centric distance is normalized relative to $R_{200m}$, their fitting functions were effectively redshift and halo mass-independent.  In our case, the dependence of the nonthermal pressure component on accretion rate is important because halo accretion in turn depends on halo concentration: lower concentration halos are expected to have higher accretion rates and thus higher nonthermal pressure support.  Since nonthermal pressure does not produce an SZ signal, this correlation leads to a suppression of the SZ signal at low halo concentration.  Explicitly, we adopt the \citet{Nelson:2014} fitting function for the  non-thermal pressure support fraction as a function of radius: 
\begin{multline}
\label{eq:fnth}
f_{\rm nth}(r) = 1 - \\
(0.509 - 0.026 \Gamma_{200m})\left( 1 + \exp\left[ -\left( \frac{r/R_{200m}}{ B}\right)^{\gamma} \right] \right),
\end{multline}
where $\Gamma_{200m} \equiv \Delta \ln (M_{200m}(a))/\Delta \ln (a)$ is the halo accretion rate. To relate $\Gamma_{200m}$ to the concentration, we perform a fit to the $\Gamma_{200m}$ and $c_{200m}$ measurements from \thethreehundred\  simulations, finding
\begin{equation}
\label{eq:gamma}
    \Gamma_{200m} \approx 5.92 c_{200m}^{-0.697}.
\end{equation}
The thermal pressure, which is responsible for the SZ signal, is then given by 
\begin{equation}
P_{\rm th}(r) = P_{\rm gas}(r)(1-f_{\rm nth}(r)).
\end{equation}

We have also tested alternative fitting functions for the nonthermal pressure support from \citet{Shaw:2010}, which do not include dependence of the nonthermal pressure on accretion rate.  In this case, as expected, we find that the SZ signal at low concentration is slightly enhanced.  The difference in our main results between using these two fitting functions is generally small, although we find that using the \citet{Shaw:2010} model results in a small (roughly 10\%) mismatch between the amplitudes of the predicted and simulated pressure profiles.  We also find that the \citet{Nelson:2014} model provides a better fit to the full $Y$-$M$-$c$ relation than the \citet{Shaw:2010} model (see \S\ref{sec:YMc}).

To account for effects such as radiative cooling and more complex nonthermal pressure support, we  
allow the amplitude of the pressure to have some additional scaling with mass:
\begin{equation}
    P_{\rm gas}(r) = (M/10^{15}\,\mathrm{M_{\odot}} )^{\eta} P_{\rm KS}(r),
\end{equation}
where $\eta$ is a free parameter, and we expect $\eta \approx 0$.

\subsubsection{Two-halo term}
\label{sec:two-halo}

At large cluster-centric distances, $r \gtrsim R_{200c}$, the contributions of nearby halos to the gas pressure become significant.  This is the so-called two-halo term \citep{Cooray:2002}.  Because the SZ effect is sensitive to the line-of-sight integrated gas pressure, two-halo contributions to the SZ signal must be considered even at projected radii $R < R_{200c}$.  In principal, the two-halo term can be modeled using the standard tools of the halo model \citep{Cooray:2002, Vikram:2017}.  Here, we will instead take the simpler approach of modeling the two-halo contributions to the pressure with a power law in radius.  Out to distances of a few $R_{200c}$, the power-law model is expected to work well.  Indeed, the power law approach can in some cases work better than standard halo model-based approaches for modeling the one-to-two halo transition regime \citep[e.g.][]{Diemer:2014}.  We discuss exactly how we connect the two-halo term to the one-halo term in more detail in \S\ref{sec:resultsII}.   In principle, gas could also be ejected out of the halo to $r > R_{200c}$ as a result of e.g. AGN feedback.  However, this effect is expected to be small for the cluster-scale halos that we consider here.

\begin{figure*}
    \centering
    \includegraphics[scale=0.4]{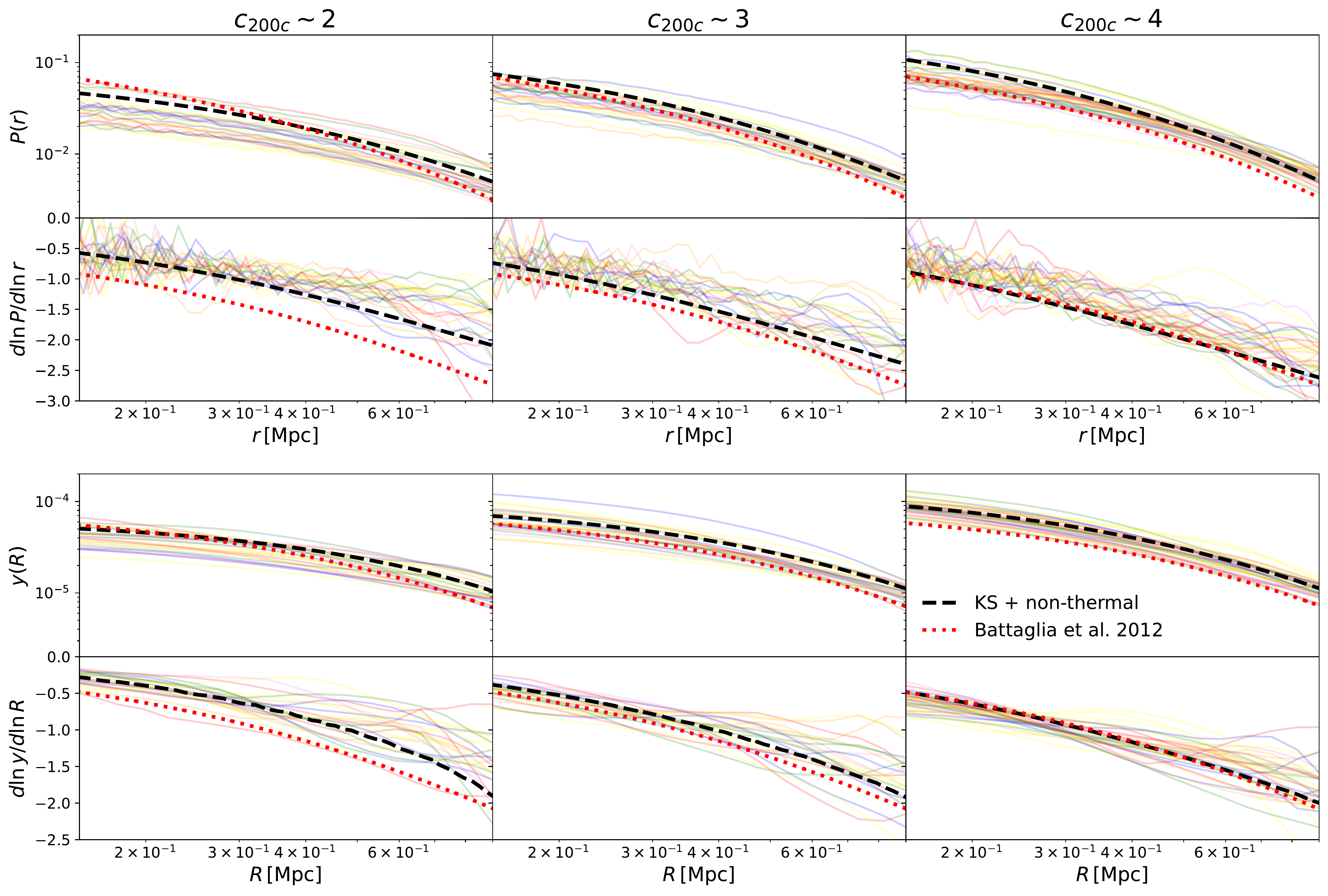}    
    \caption{The inner pressure profiles (top panels) and SZ profiles (bottom panels) for halos of mass $8\times 10^{14} < M_{200c}/\mathrm{M_{\odot}} < 1.2 \times 10^{15}$ 
    and different concentration (see headings at top of columns).  The quantity $r$ represents the 3D distance from the cluster center, while $R$ represents the projected radial distance from the line of sight.  The thin solid curves represent individual halos from The300 simulations.  The black dashed curve corresponds to the \citetalias{KomatsuSeljak2001} model with the addition of non-thermal pressure support, as described in \S\ref{sec:polytrope}.  The \citetalias{KomatsuSeljak2001} model captures the dependence on concentration of the inner pressure and SZ profiles seen in the simulations.  The red dotted curve corresponds to the fitting formula from \citet{Battaglia:2012}, which does not include dependence on concentration (these curves are the same in each row).}
    \label{fig:profile_smallR}
\end{figure*}

\subsubsection{Conversion to SZ profile}

The SZ profile, $y(R)$, at projected distance from the cluster, $R$, is 
\begin{equation}
\label{eq:pressure_to_sz}
    y(R) = \frac{\sigma_T}{m_e c^2} \int dl P_e(r = \sqrt{R^2 + l^2}), 
\end{equation}
where $P_e(r)$ is the thermal electron pressure and $l$ is the line of sight distance.  We relate the electron pressure to the total thermal pressure using
\begin{equation}
P_e(r) \approx \frac{4 - 2Y}{8-5Y} P_{\rm th}(r) \approx 0.518 P_{\rm th}(r),
\end{equation}
where $Y \approx 0.24$ is the primordial Helium mass fraction, and we have assumed that the Hydrogen and Helium are fully ionized.  In principle, the integral in Eq.~\ref{eq:pressure_to_sz} should be over the entire line of sight to the last scattering surface.  However, since the cluster environment dominates the integral, we restrict the limits of integration to $\pm 5 R_{200c}$.  We find that extending the integration limits does not significantly impact our results.   
The integrated SZ signal is then given by 
\begin{equation}
\label{eq:integratedY}
Y_{\Delta} = \int_{0}^{R_{\Delta}} dR \,y(R) 2 \pi R .
\end{equation}

\section{Results I: Impact of concentration on the inner SZ profile}
\label{sec:resultsI}

We first consider how concentration impacts the inner SZ profiles at $R \lesssim 1\, {\rm Mpc}$.  Fig.~\ref{fig:profile_smallR} compares the 3D profiles  (top two rows) of individual simulated clusters from the Three Hundred Project to the \citetalias{KomatsuSeljak2001} model with non-thermal pressure support (black dashed curves) in bins of concentration and at cluster mass $M \sim 10^{15}\,\mathrm{M_{\odot}}$.  As concentration is increased, the inner pressure profiles of the simulated clusters increase in amplitude and become steeper, behavior which is predicted by the \citetalias{KomatsuSeljak2001} model as described above.  

The second row of Fig.~\ref{fig:profile_smallR} shows the logarithmic derivative of the pressure profiles.  To reduce scatter in the derivatives resulting from small variations in gas density, we apply a Savitzky–Golay filter to the profile measurements before taking the derivative.  For the 3D pressure profile measurements (which use 150 bins), we use a window length of 15 and a polynomial order of three.  The \citetalias{KomatsuSeljak2001} model (with the inclusion of non-thermal pressure support) shows good agreement with the logarithmic derivatives measured from the simulations.  For comparison, we also show the fitting function from \citet{Battaglia:2012}, which does not include variation in the pressure profiles with concentration.  It is clear that this function describes the profiles well at $c_{200c} \gtrsim 3$, but overpredicts the amplitude and steepness of the pressure profile at low $c_{200c}$. 

\begin{figure*}
    \centering
    \includegraphics[scale=0.4]{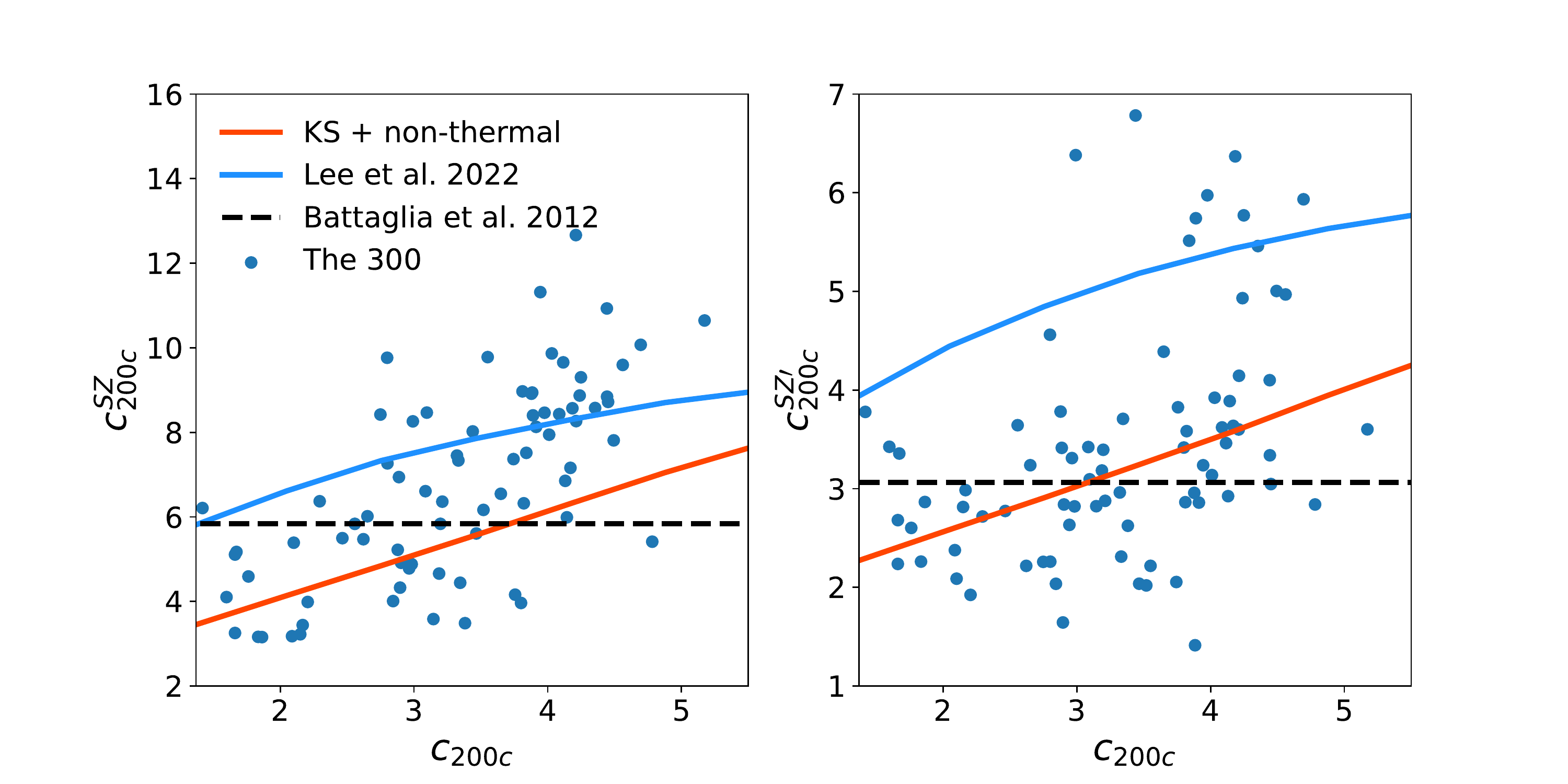}    
    \caption{The relation between SZ concentration, $c^{SZ}_{200c}$, (defined in Eq.~\ref{eq:sz_concentration}) and the halo concentration, $c_{200c}$, for halos with $9 < M/(10^{14}\,\mathrm{M_{\odot}}) < 11$.  The left panel shows the result when defining the SZ scale radius based on where the logarithmic slope reaches -1, probing the deep interior of the cluster, while the right panel defines the scale radius based on where the logarithmic slope reaches -1.5, thus probing farther into the cluster outskirts.   The red curve shows the prediction of the \citetalias{KomatsuSeljak2001} model (with non-thermal pressure support), while the blue curve represents the predictions of \citet{Lee:2022}.  The black dashed curve indicates the prediction from the \citet{Battaglia:2012} model, which does not include concentration dependence. 
  }
    \label{fig:SZc_vs_massc}
\end{figure*}

The bottom two rows of Fig.~\ref{fig:profile_smallR} show the SZ profiles of the same clusters.  Since we only use 30 radial bins for measuring the SZ profiles, there is somewhat less scatter in these profiles and we use a Savitzky-Golay window size of five and polynomial order of three when computing the logarithmic derivative.  Again, we see that the \citetalias{KomatsuSeljak2001} model does an excellent job of capturing the trends in the SZ profiles with varying concentration.  

To aid comparison between the model predictions and the simulated clusters, we define an SZ analog of the halo concentration parameter:
\begin{equation}
\label{eq:sz_concentration}
c^{\rm SZ}_{\Delta} \equiv R_{\Delta}/R_{s}^{\rm SZ},
\end{equation}
where $R^{\rm SZ}_{s}$ is a scale radius for the SZ profile.  The scale radius for the dark matter density profile is the location where its logarithmic slope reaches -2.  Since the SZ profile is obtained from the 3D pressure profile by integrating along the line of sight, we define $R^{\rm SZ}_s$ as the radius where the SZ profile reaches a logarithmic slope of -1.  In order to test the profile predictions at slightly larger radius, we also define a modified SZ concentration, ${c^{\rm SZ}_{\Delta}}' = R_{\Delta}/{R_s^{\rm SZ}}'$, where ${R_s^{\rm SZ}}'$ is the radius where the logarithmic slope of the SZ profile reaches -1.5.

Fig.~\ref{fig:SZc_vs_massc} shows the relation between $c_{200c}$ and $c_{200c}^{\rm SZ}$ in the simulations (points) compared to the predictions of the \citetalias{KomatsuSeljak2001} model with non-thermal pressure support (red curve).   For comparison, we also show the prediction of the fitting function from \citet{Lee:2022} as well as the prediction of \citet{Battaglia:2012} (which does not include variation in the profile with concentration).  For both $c_{200c}^{\rm SZ}$ and ${c^{\rm SZ'}_{200c}}$, the \citetalias{KomatsuSeljak2001} model reproduces the trends reasonably well.  Note the high values of $c_{200c}^{\rm SZ}$, with $c_{200c}^{\rm SZ} \gtrsim 8$ in some cases.  This means that the region of the SZ profile being probed is at very small scales (less than $\approx R_{\Delta}/8$).  It is perhaps not surprising, then, that the \citetalias{KomatsuSeljak2001} model does not perfectly capture the trends seen at high $c_{200c}^{\rm SZ}$.  At small scales, the impact of e.g. cooling and feedback may be significant, which are not captured by the polytropic model.  Feedback and cooling effectively introduce new scales into the problem (for instance, the radius at to which gas is ejected by AGN, \citealt{Pandey:2023}), breaking the assumption of self-similarity in the \citetalias{KomatsuSeljak2001} model.  We see that ${c^{\rm SZ}_{\Delta}}'$, which effectively probes the pressure profile at larger radii, is better described by the polytropic model.   

There are several halos that have very high ${c^{\rm SZ}_{\Delta}}$ and ${c^{\rm SZ}_{\Delta}}'$  which lie far from the relation predicted by the \citetalias{KomatsuSeljak2001} model.  We find that these halos have very steep inner pressure profiles (explaining their large values of ${c^{\rm SZ}_{\Delta}}$), and relative to other halos of the same mass, they have higher amplitude inner pressure profiles, lower amplitude inner entropy profiles, and higher inner stellar density profiles.  These observations are consistent with the high-${c^{\rm SZ}_{\Delta}}$ clusters being cool core clusters \citep{Hudson:2010}.  

\section{Results II: Impact of concentration on the outer SZ profile and the $Y$-concentration relation}
\label{sec:resultsII}

\subsection{Comparing the \citetalias{KomatsuSeljak2001} model to simulations in the halo outskirts}

We now consider the outer pressure profiles and the relation between $Y_{\Delta}$ and $c_{\Delta}$.  The quantity $Y_{\Delta}$ is often the SZ observable used in cluster cosmology analyses as it is (effectively) directly observable and correlates tightly with cluster mass.  Some SZ-cluster analyses use the SZ signal-to-noise instead \citep[e.g.][]{Bocquet:2019}, but this quantity is also closely related to $Y_{\Delta}$.  We will consider common apertures choices of $R_{500c}$ and $R_{200c}$ for defining $Y_{\Delta}$.   

\begin{figure*}
    \centering
    \includegraphics[scale=0.5]{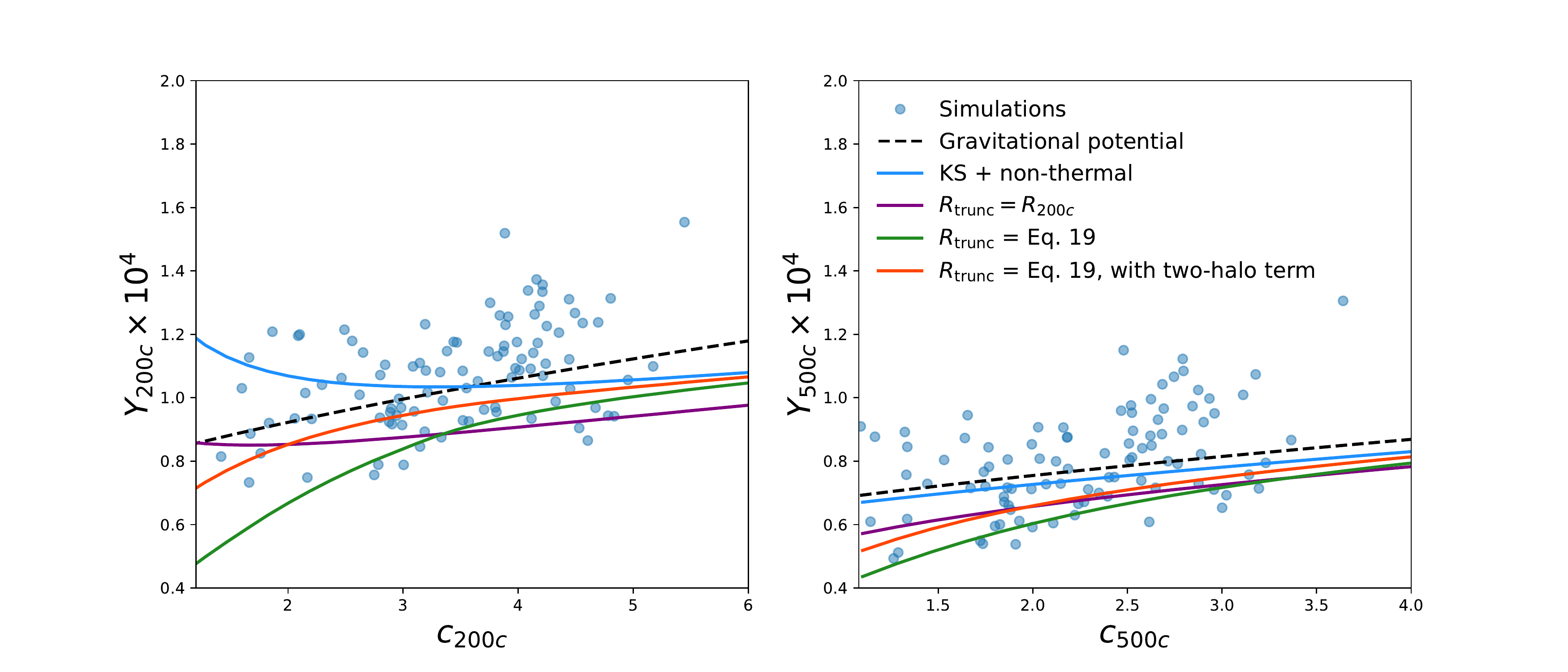}
    \caption{The integrated SZ signal within $R_{200c}$ (left) and $R_{500c}$ (right) for clusters from The300 as a function of the corresponding concentration parameter, restricted to clusters with $8.5 < M/(10^{14} \,\mathrm{M_{\odot}}) < 11.5$ (blue points).  Without truncation of the SZ profile, the \citetalias{KomatsuSeljak2001} model with non-thermal pressure support (blue curve) significantly overpredicts the integrated signal at low concentration.  Truncating the pressure profile at $R_{200c}$ (purple curve) improves the agreement somewhat, but is not sufficient to fully explain the $Y_{200c}$-$c_{200c}$ relation.  Varying the truncation radius according to Eq.~\ref{eq:R_t} (green curve) yields better agreement.  Combining this model with a prescription for the two-halo term from Eq.~\ref{eq:final_model} (red curve) yields good agreement with the simulated $Y_{\Delta}$-$c_{\Delta}$ relation.}  
    \label{fig:Y_vs_concentration}
\end{figure*}

Fig.~\ref{fig:Y_vs_concentration} shows the relation between $Y_{200c}$ ($Y_{500c}$) and $c_{200c}$ ($c_{500c}$) in the left (right) panel from The300, restricting to clusters with $M\sim 10^{15}\,\mathrm{M_{\odot}}$ (blue points).  There is a trend for clusters with higher concentration to have higher $Y_{200c}$, as noted previously in Fig.~\ref{fig:YM_scatter}.  The black dashed curve shows the predicted $Y_{\Delta}$-$c_{\Delta}$ relation based on the gravitational potential energy (\S\ref{sec:potential_argument}), scaled in amplitude to best match the simulation results.  It appears that this curve does a good job of reproducing the shape of the $Y$-$c$ relation.  

The blue solid curves in Fig.~\ref{fig:Y_vs_concentration} represent the predicted relation between $Y_{\Delta}$ and $c_{\Delta}$ from the \citetalias{KomatsuSeljak2001} model with non-thermal pressure support (\S\ref{sec:polytrope}), not including the two-halo contribution.  As in \S\ref{sec:resultsI}, we have adopted $\eta = 0$; note that the value of $\eta$ is inconsequential here since we are considering fixed cluster mass.  It is clear from the left panel of Fig.~\ref{fig:Y_vs_concentration} that this model fails to capture the measured $Y_{200c}$-$c_{200c}$ relation: at low concentration, the model predicts a steep rise in the integrated SZ signal that is not seen in the simulations.  Including the two-halo contribution here could only increase $Y$, and would therefore not resolve this discrepancy.  This model provides a better description of the $Y_{500c}$-$c_{500c}$ relation (right panel of Fig.~\ref{fig:Y_vs_concentration}), which receives contributions from smaller cluster-centric radii. 

The origin of the increase in $Y_{\Delta}$ predicted by the polytropic model at low $c_{\Delta}$ can be traced to the outer pressure profiles.  As seen in Fig.~\ref{fig:polytrope}, the \citetalias{KomatsuSeljak2001} model predicts that as concentration is decreased, the pressure rises in the cluster outskirts.  Since the cluster outskirts contribute significantly to the volume integral of the pressure, the increase in pressure is sufficient to drive an increase in $Y_{\Delta}$ when the concentration is low ($c_{200c} \lesssim 4$).  This effect is more pronounced for the $Y_{200c}$ vs. $c_{200c}$ relation than for the $Y_{500c}$ vs. $c_{500c}$ relation because $Y_{200c}$ receives more contributions from the cluster outskirts.  Note that because the SZ signal is related to an integral along the line of sight, $Y_{500c}$ receives contributions from $r > R_{500c}$.  

In order to explain the $Y$-$c$ relation measured in simulations, there must be some rapid steepening of the outer pressure profile to reduce its contribution to $Y$ at low concentration.  Indeed, a rapid decline in the pressure profile at the location of the virial shock is predicted by self-similar collapse models, as we discuss in more detail in \S\ref{sec:shocks}.  As a simple model for this rapid steepening, we multiply the cluster pressure profile by a top-hat function centered at radius $R_{\rm trunc}$:
\begin{equation}
f_{\rm trunc}(r; R_{\rm trunc}) = \Theta(R_{\rm trunc} - r),
\end{equation}
where $\Theta$ is a Heaviside function.  The result of setting $R_{\rm trunc} = R_{200c}$ is shown with the purple curve in Fig.~\ref{fig:Y_vs_concentration}.  It is clear that this truncation model reduces $Y$ at low concentration as desired, but now $Y_{200c}$ is under-predicted at high concentration, and we still see an increase in $Y_{200c}$ at low concentration which does not match the simulations.  Note again that inclusion of the two-halo term could only increase $Y$ and would not resolve the discrepancy at low concentration.  

\begin{figure}
    \centering
    \includegraphics[scale=0.45]{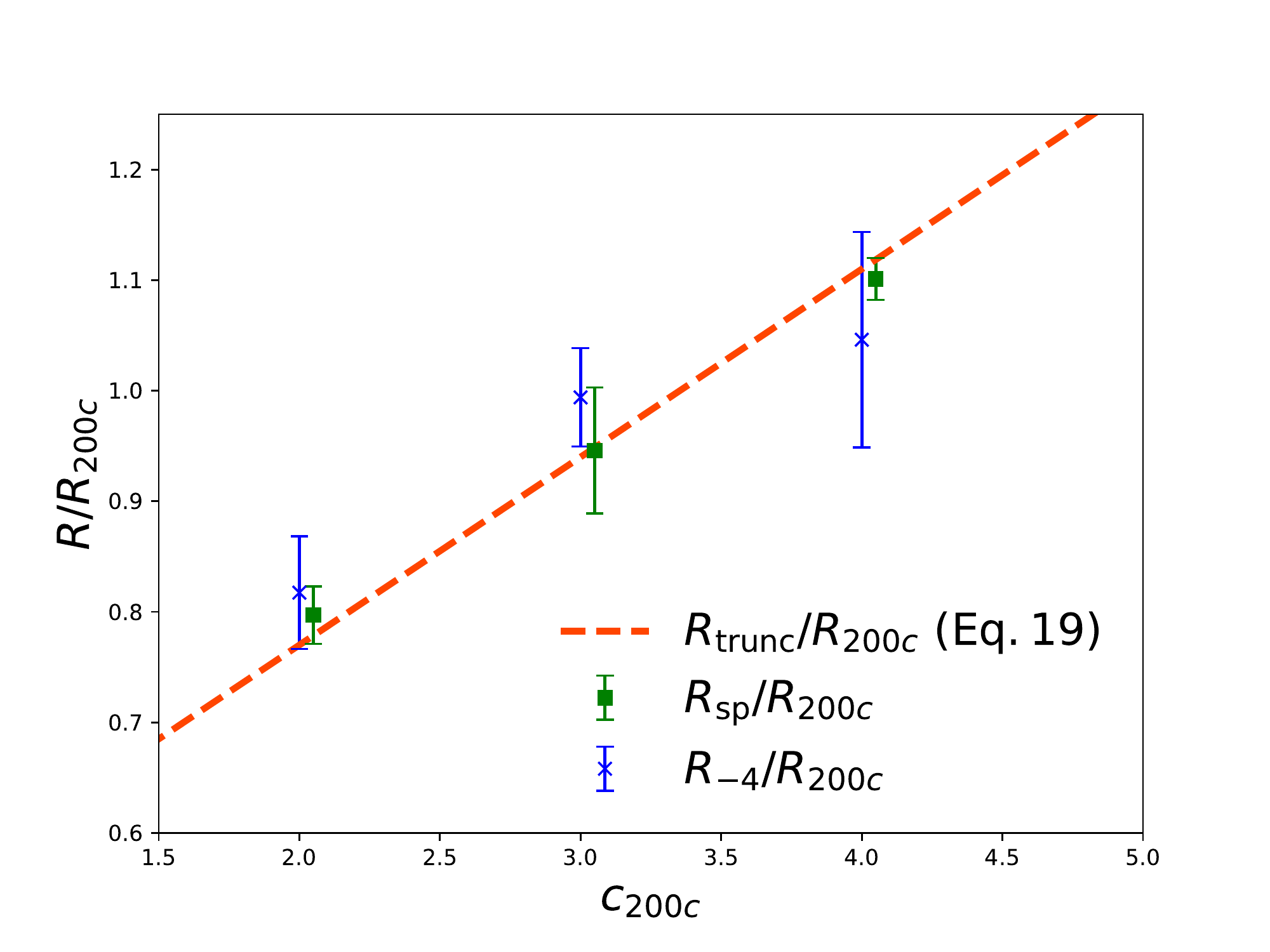}
    \caption{The gas truncation radius $R_{-4}/R_{200c}$ as a function of concentration (blue points).  The red curve shows the truncation radius model from Eq.~\ref{eq:R_t}.  The green points represent the splashback radius measured from the density profiles for clusters from The300 with $0.8 \times 10^{15} < M /\mathrm{M_{\odot}} < 1.2 \times 10^{15}$ and concentration in three bins centered at $c_{200c} = 2,3,4$.    }
    \label{fig:Rsp_and_Rtruncation}
\end{figure}

It appears, then, that in order to match the observed $Y_{\Delta}$-$c_{\Delta}$ relation, $R_{\rm trunc}$ must vary with concentration.  As a simple starting point, we allow the truncation radius to vary linearly with concentration as
\begin{equation}
\label{eq:R_t}
    R_{\rm trunc}/R_{200c} = a + bc_{200c},
\end{equation}
where $a$ and $b$ are free parameters.  In order to constrain the values of $a$ and $b$, we first determine the approximate location of the steepening of the pressure profiles by identifying the radii at which the logarithmic slopes of these profiles reach $-4$, which we refer to as $R_{-4}$.  As expected, we find that $R_{-4}$ increases with increasing concentration (Fig.~\ref{fig:Rsp_and_Rtruncation}).  We find that setting $(a,b) = (0.43, 0.17)$ provides a good match to this trend.  The predicted $Y$-$c$ relation for this choice of parameters is shown with the green curve in Fig.~\ref{fig:Y_vs_concentration}.  We see that by decreasing the truncation radius as concentration is decreased, the predicted $Y_{\Delta}$ now decreases monotonically as concentration is decreased, in agreement with the trend seen in simulations.

\begin{figure*}
    \centering
    \includegraphics[scale=0.4]{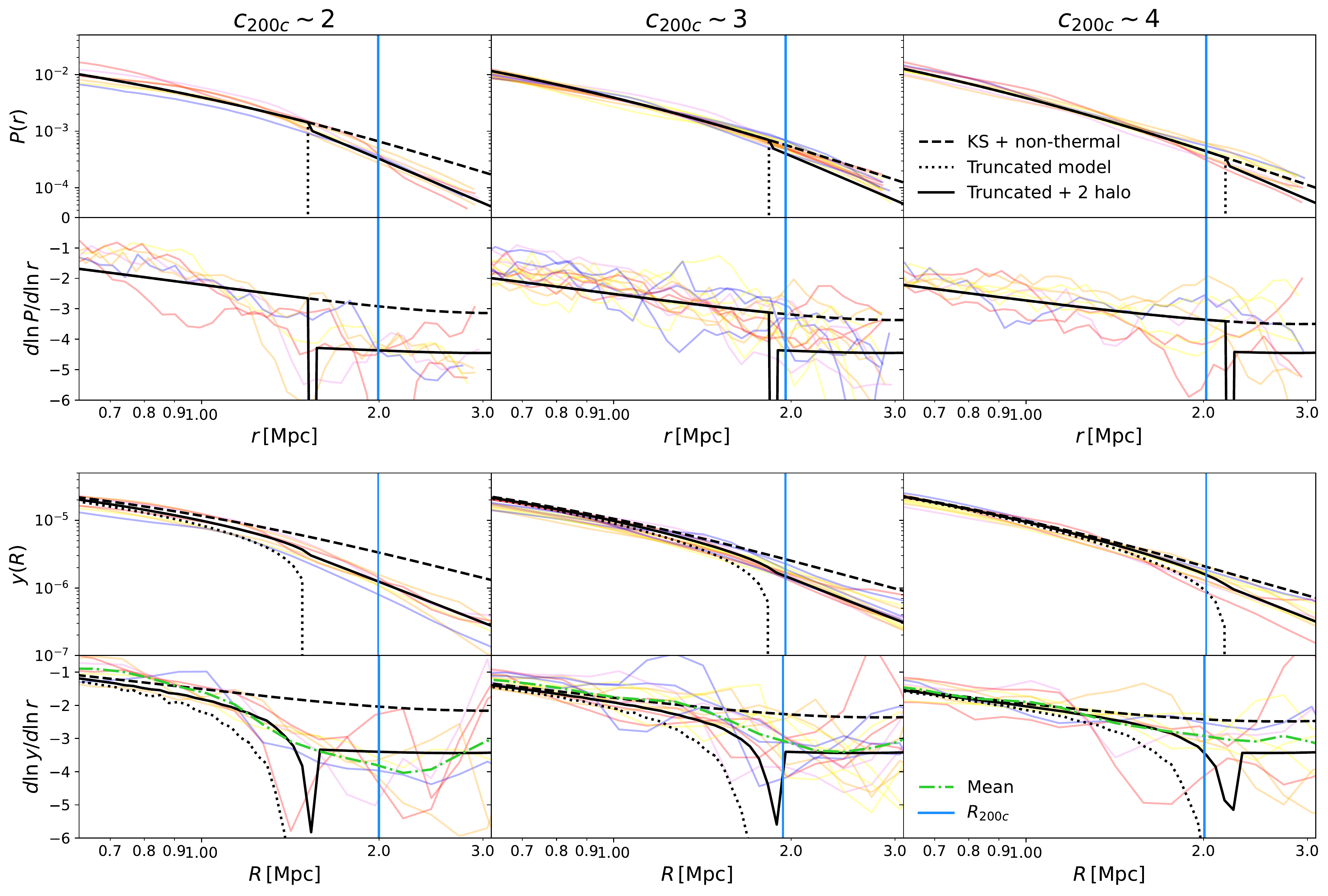}    
    \caption{The outer pressure (top panels) and SZ profiles (bottom panels) for clusters of different concentrations (three columns).  The black curves represent different variations of the polytrope model: the \citetalias{KomatsuSeljak2001} model with non-thermal pressure support (dashed curve), the truncated \citetalias{KomatsuSeljak2001} model with non-thermal pressure support and truncation radius given by Eq.~\ref{eq:R_t} (dotted curve), and the same with the addition of the two-halo term (Eq.~\ref{eq:final_model}, solid curve).   The latter provides a good match to the pressure and $y$ profiles across a range of concentration. The blue vertical lines indicates $R_{200c}$, while the green dot-dashed curves in the bottom row show the mean logarithmic slope profile across the clusters in the bin.  The faded colored curves represent the profiles of individual simulated clusters.}
    \label{fig:profile_largeR}
\end{figure*}

Finally, we must add the two-halo term to the model calculation.  As described in \S\ref{sec:two-halo}, we treat the two-halo term as a power law in cluster-centric distance.  We set the amplitude of the two-halo term to be a fixed fraction, $f$, of the cluster-pressure profile at the truncation radius.  In other words, we set
\begin{equation}
    P_{2h}(r) = f P_{\rm th}(R_{\rm trunc})\left(r/R_{\rm trunc} \right)^{\alpha} \Theta(r-R_{\rm trunc}),
\end{equation}
where $f$ and $\alpha$ are parameters of the model, and we set $P_{2h}(r) = 0$ for $r < R_{\rm trunc}$.  We fix $\alpha = -4$ and $f = 0.75$, as these values provides a good match to the outer pressure profiles of the simulated clusters (see Fig.~\ref{fig:profile_largeR}).  We note, though, that our model is not very sensitive to variations in the value of $f$ or $\alpha$.  Even setting $f=1$ provides a reasonable match to the $Y_{\Delta}$-$c_{\Delta}$ relation, although it does not provide quite as good agreement to the 3D pressure profiles.  Our intent here is not to fine tune the parameters to best match The300, since our model is merely a qualitative description of the truncation of the profiles anyways.  Rather, we aim to find a simple and physically motivated description of the pressure profiles that reasonably matches the behavior seen in the simulations.   The inclusion of the two-halo term (red curve in Fig.~\ref{fig:Y_vs_concentration}) increases the amplitude of $Y$ at low concentration, and the resultant model exhibits excellent agreement with the measurements from The300 simulation. 

To summarize, our final model for the thermal pressure profile is 
\begin{equation}
\label{eq:final_model}
    P_{\rm th}(r) = \left(\frac{M_{200c}}{10^{15}\,\mathrm{M_{\odot}}}\right)^{\eta} \left[P_{\rm inner}(r) + P_{\rm outer}(r)  \right] 
\end{equation}
where 
\begin{equation}
    P_{\rm inner}(r) = P_{\rm KS}(r) (1-f_{\rm nth}(r))\Theta(R_{\rm trunc} - r) 
\end{equation}
and 
\begin{multline}
    P_{\rm outer}(r) = f P_{\rm KS}(R_{\rm trunc}) (1-f_{\rm nth}(R_{\rm trunc}))\\
    \Theta(r - R_{\rm trunc})\left(\frac{r}{R_{\rm trunc}} \right)^{\alpha},
\end{multline}
with $P_{\rm KS}(r)$ given by the polytropic model from \citetalias{KomatsuSeljak2001} (with normalization set by the requirement that the integrated baryon density match the cosmic mean), $f_{\rm nth}(r)$ given by Eq.~\ref{eq:fnth}, and $R_{\rm trunc}$ given by Eq.~\ref{eq:R_t} with $a = 0.43$, $b = 0.17$.  We fix $\alpha = -4$ and $f = 0.75$.  In this section, since we consider fixed $M_{200c} = 10^{15}\,M_{\odot}$, the value of $\eta$ is unimportant.  In \S\ref{sec:YMc},  when we consider a wider range of $M_{200c}$, we will fit for $\eta$.   

We show the corresponding pressure (top two panels) and $y$ (bottom two panels) profiles in Fig.~\ref{fig:profile_largeR},  now extending the plot to larger $R$ than in Fig.~\ref{fig:profile_smallR}.  The blue vertical lines in the figure illustrate the mean $R_{200c}$ for reference.  It is clear why the \citetalias{KomatsuSeljak2001} model without truncation leads to larger than expected $Y$ at low concentration: this model shows a significant excess in the pressure and $y$ profiles around and beyond $R_{200c}$, particularly at low concentration.  It is also clear that the radius of the departure from the \citetalias{KomatsuSeljak2001} profile is not a fixed fraction of $R_{200c}$: at low concentration, the profile steepens at $r < R_{200c}$, while at high concentration, the profile steepens at $r > R_{200c}$.  This is consistent with the fact that our truncation radius model (Eq.~\ref{eq:R_t}) has $R_{\rm trunc}$ increase with increasing concentration.

Of course, one would not expect the \citetalias{KomatsuSeljak2001} model to  perform very well outside $R_{200c}$ since this model does not include two-halo contributions.  However, the model we have introduced above additionally describes the $y$ profiles better than \citetalias{KomatsuSeljak2001} within $R_{200c}$.  This is because the SZ profile at projected radius $R$ is sensitive to the pressure profile at scales $r > R$.  The green dot-dashed curves in the bottom row of Fig.~\ref{fig:profile_largeR} show the mean log-derivative $y$ profile across the clusters in each concentration bin, which agrees very well with our simple model.  The model predicts a sharp steepening of the profile near the truncation radius, as expected since we have assumed maximally steep truncation of the pressure profile.  This sharp truncation is not seen in the simulations, as to be expected for several reasons.  For one,    effects like asphericity of the pressure profiles of individual halos will lead to smoothing of the truncation regime, since the truncation will not occur at a single radius.  Additionally, averaging across multiple clusters with different truncation radii will necessarily decrease the sharpness of the truncation.  Indeed, many individual clusters show regions of very steep decline in their $y$ profiles, with logarithmic slopes below $-4$, while the averaged $y$ profile never becomes this steep.  We emphasize that it is not our intent here to perform a detailed modeling of the pressure profiles near the truncation region.  As we show below, even the simple model that we have adopted is sufficient to yield significantly improved predictions for the $Y$-concentration relation relative to the \citetalias{KomatsuSeljak2001} model.

\subsection{Physical origin of the truncation}
\label{sec:shocks}

The finding that the pressure profiles should be truncated near $R_{200c}$ is expected given the predictions of secondary infall models \citep{Bertschinger:1985}.  \citet{Bertschinger:1985} found that in such models, the gas undergoes a shock as it accretes onto the cluster.   Gas passing through the shock experiences a rapid increase in pressure.  The location of the shock coincides roughly with the location of first apocenter of accreting dark matter \citep{Bertschinger:1985, Shi:2016}, now known as the splashback radius \citep{Diemer:2014, AD14}.  The splashback radius in turn anti-correlates with accretion rate \citep{Diemer:2014, AD14, Shi:2016}, and thus correlates positively with concentration.  Our finding that the truncation radius must decrease with decreasing concentration in order to explain the $Y_{\Delta}$-$c_{\Delta}$ relation is consistent with this picture.

To investigate the relation between the gas truncation radius and the splashback radius, we measure the splashback radius for the clusters from The300 in the same mass and concentration bins as in Fig.~\ref{fig:profile_largeR}.  To do this, we calculate the stacked dark matter density profile for the clusters in the three concentration bins and find their logarithmic slope using the Savitzky-Golay method \citep{Diemer:2014}.  The splashback radius is then given by the radius at which the logarithmic slope is minimized. The error bars are estimated using a jackknife sampling method.

Fig.~\ref{fig:Rsp_and_Rtruncation} shows a comparison of the splashback measurements (green points) compared to our model for the pressure profile truncation radius, $R_{\rm trunc}$ (red dashed curve), from Eq.~\ref{eq:R_t}.  It is clear that the truncation radius model tracks $R_{\rm sp}$ very closely as a function of concentration.   We also plot the radii at which the logarithmic derivatives of the 3D pressure profiles reach a slope of -4 (blue points with errorbars).   Our model for the truncation radius roughly coincides with this location.  

Previous work has also found evidence for signatures of shocks in the SZ profiles of simulated galaxy clusters near to the halo splashback radius.  \citet{Molnar:2009} identified what they referred to as ``virial shocks'' at $0.9$ to $1.3 R_{200m}$ in the SZ and pressure profiles of simulated galaxy clusters.  Our finding that the pressure profile truncation radius coincides with the splashback radius is consistent with their results.  

We note, though, that the shock located at the splashback radius is not the only shock front expected around galaxy clusters.  For instance, \citet{Molnar:2009} identified ``external'' shocks at several virial radii that also lead to features in the SZ profiles of galaxy clusters.  Such signatures were also investigated by \citet{Baxter:2021}.  \citet{Zhang2022} have shown that such external shocks may result from mergers driving shock fronts to large radii (i.e. so-called merger accelerated accretion shocks).  Shocks at these very large distances from the cluster center are unlikely to have a significant impact on the $Y$-$c$ relation.

\begin{figure*}
    \centering
    \includegraphics[scale = 0.5]{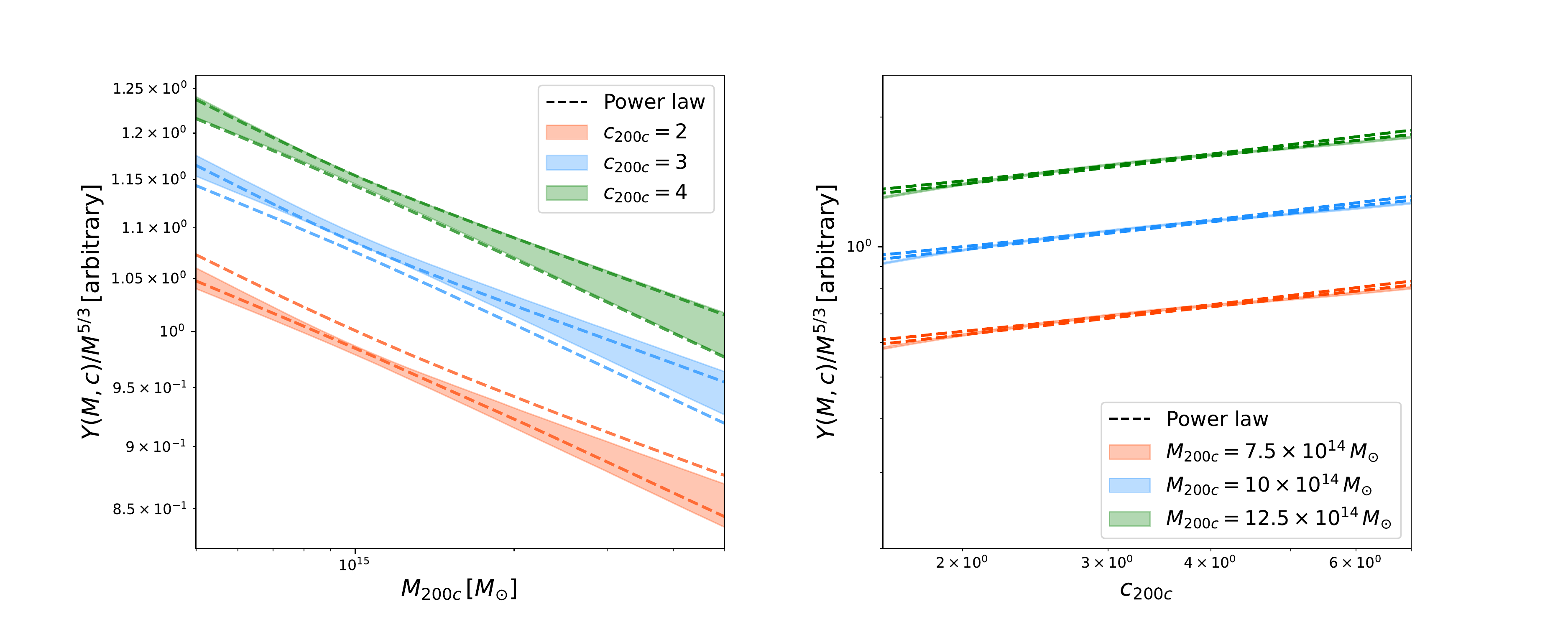}
    \caption{Inferred relationship between $Y_{200c}$, $M_{200c}$, and $c_{200c}$ as a function of mass (left) and concentration (right).  The filled bands indicate the 68\% constraints obtained using the model of Eq.~\ref{eq:final_model}, while the dashed lines indicate the constraints from the power law model.} 
    \label{fig:fit_results}
\end{figure*}

\section{Results III: The $Y$-$M$-$c$ relation and its implications}
\label{sec:resultsIII}

\subsection{Fitting the $Y$-$M$-$c$ relation}
\label{sec:YMc}

We have argued above that the truncated pressure profile model introduced in Eq.~\ref{eq:final_model} provides a better description of the pressure and SZ profiles of clusters than the standard \citetalias{KomatsuSeljak2001} model, and also that it leads to significantly improved predictions  for the $Y$-$c$ relation at fixed cluster mass.  We now assess whether this model provides a good description of the full $Y$-$M$-$c$ relation and determine the best choice of the parameter $\eta$.  To do this, we introduce additional amplitude freedom into our pressure profile model, rescaling $P_{\rm th}(r) \rightarrow A_M P_{\rm th}(r)$.  We then fit the $Y$ measurements from \thethreehundred\, keeping $A_M$ and $\eta$ free.  Recovering $A_M \approx 1$ will provide a test of our model. To perform this fit, we assume log-normal scatter in $Y_{\Delta}$ around the mean $Y$-$M$-$c$ relation:
\begin{dmath}
\label{eq:Y_M_c}
    P(\ln Y^{\rm sim}_{\Delta} | M,c) = \\ \frac{1}{\sqrt{2\pi \sigma_{\ln Y_{\Delta}}^2}}\exp\left[-\frac{(\ln Y^{\rm sim}_{\Delta} - \ln Y_{\Delta}(M,c))^2}{2\sigma_{\ln Y_{\Delta}}^2} \right],
\end{dmath}
where we treat $\sigma_{\ln Y}$ as a free parameter and 
$Y_{\Delta}$ is given by Eq.~\ref{eq:integratedY} and Eq.~\ref{eq:final_model}.  We fix the parameters $f$, $a$, $b$ and $\alpha$ to the values quoted in the text.  The remaining free parameters are $\vec{\theta} = \left( A_M, \eta, \sigma_{\ln Y} \right)$.

For the purposes of model comparison, we also consider a pure-power law relation between $Y$-$M$-$c$:
\begin{equation}
    Y(M,c) = A_p M^{\eta_p} c^{\beta_p},
\end{equation}
treating $\vec{\theta}_{p} = \{A_p, \eta_p, \beta_p\}$ as free parameters.  We expect $\eta_p \approx 5/3$ in accordance with the self-similar prediction.

We compute the posterior on the model parameters, $\vec{\theta}$ and $\vec{\theta}_p$, via
\begin{equation}
    P(\vec{\theta} | \{ Y^{\rm sim}_i, M^{\rm sim}_i, c^{\rm sim}_i\}) = \prod_i^{N_c}  P_Y(\ln Y^{\rm sim}_i | M^{\rm sim}_i,c^{\rm sim}_i, \vec{\theta}),
\end{equation}
where we have assumed flat priors on the model parameters, and we treat each cluster as independent, taking the product across all $N_c$ clusters.  Since our focus is on massive clusters, we restrict our fits to those clusters with $M_{200c} > 5\times 10^{14}\,\mathrm{M_{\odot}}$.

The results of our truncated pressure profile model and power-law fits to The300 measurements are shown in Fig.~\ref{fig:fit_results}. We find $A_M = 0.98\pm0.01$ and $\eta = -0.09\pm 0.02$.  Reassuringly, we find that $A_m \approx 1$, suggesting that our nonthermal pressure support model is providing a good description of \thethreehundred\ simulations. The shaded bands represent the 68\% credible intervals on the $Y$-$M$ (left) and $Y$-$c$ (right) relations using the modified \citetalias{KomatsuSeljak2001} model, while the dashed curves indicate the same for the power law model.  The results of the two fits look generally similar, indicating that the predicted $Y_{\Delta}$-$c_{\Delta}$ relation does indeed provide a good description of the simulations across a range of cluster masses and concentrations.  Note, though, that the truncated pressure profile model introduced here predicts a departure from a pure power law in the $Y$-$c$ relation at high and low concentration.  Both models yield very similar levels of scatter: $\sigma_{\ln Y} = 0.099 \pm 0.004$ for the model from Eq.~\ref{eq:final_model}, and $\sigma_{\ln Y} = 0.098 \pm 0.004$ for the power law model.  This level of scatter is consistent with previous studies in simulations \citep[e.g.][]{Nagai:2006, Battaglia:2012, Planelles:2017} and data \citep[e.g.][]{Sifon:2013,Dietrich:2019, Sayers:2023}.

\subsection{Implications of the $Y$-$M$-$c$ relation for analyses of SZ-selected clusters}

The correlation between $Y_{\Delta}$ and $c_{\Delta}$ will result in the concentration distribution of SZ-selected cluster samples being modified: since $Y_{\Delta}$ and $c_{\Delta}$ are positively correlated, clusters with higher $c_{\Delta}$ are more likely to be detected.  This will in turn have implications for the inference of quantities that are also correlated with concentration from these samples.  We focus on two such quantities here: (1) the linear halo bias, and (2) the splashback radius.  The implications of the $Y$-$c$ relation for the former have been previously considered in \citet{Wu:2008}.  

We first determine how the $Y$-$M$-$c$ relation changes the concentration distribution for an SZ-selected sample.  We consider a cluster sample selected based on $Y > Y_{\rm min}$, where $Y_{\rm min}$ is some chosen threshold.  The concentration distribution of the resultant sample can be expressed as:
\begin{equation}
\label{eq:c_distrib}
    P(c) = \int_{Y_{\rm min}}^{\infty} dY \int dM P(Y | M,c) P(M,c),
\end{equation}
where $P(Y | M,c)$ is given by our fit the to the $Y$-$M$-$c$ relation from Eq.~\ref{eq:Y_M_c}, and $P(M,c)$ is the probability of a halo having mass $M$ and concentration $c$.   To model $P(M,c)$ we use
\begin{equation}
    P(M,c) = P(c | M)P(M),
\end{equation}
where $P(M)$ is proportional to the halo mass function; we use the mass function from \citet{Tinker:2008} in our calculations.  $P(c|M)$ encapsulates the mass-concentration relation; we model the expectation value of this relation, $\langle c | M \rangle$, using fitting formulae from \cite{Diemer:2019}.     We assume log-normal scatter around this mean relation with scatter of $\sigma(\log_{10} c) = 0.14$  \citep{Wechsler:2006}.  We compare the result of Eq.~\ref{eq:c_distrib} to a calculation that ignores the relation between $Y$ and concentration, which we refer to as $P_{\rm noc}(c_{200c})$.  To calculate $P_{\rm noc}(c_{200c})$, we replace $Y(M,c)$ in Eq.~\ref{eq:Y_M_c} with the expectation marginalized over $c$:
\begin{equation}
    Y(M) = \int dc \,Y(M,c)P(c|M).
\end{equation}
The change in the concentration distribution of the SZ-selected sample as a result of ignoring the concentration dependence of $Y$ (i.e. $P(c_{200c})/P_{\rm noc}(c_{200c})$) is shown in Fig.~\ref{fig:conc_distribution} for different values of $Y_{\rm min}$.  The concentration dependence of $Y$ causes $P(c_{\Delta})$ to shift to larger $c_{\Delta}$ since clusters with larger $c_{\Delta}$ are more likely to have $Y > Y_{\rm min}$.  We now consider how these changes to the concentration distribution impact inference of the linear bias parameter and the splashback radius.

\subsubsection{Assembly bias and cluster clustering}
\label{sec:clustering}

The clustering of clusters carries information about their masses, with higher mass clusters being more highly clustered.  In the two-halo regime, the amplitude of the clustering is proportional to  $b^2(M)$, where $b(M)$ is the linear bias parameter, which is a function of the halo mass $M$ (and redshift) that can be predicted using theory and simulations \citep[e.g.][]{Tinker:2010}.  A measurement of the clustering amplitude can then be used to infer $b$, which can be translated into a constraint on $M$.  The possibility of using clustering to calibrate cluster masses has been considered in several works \citep{Lima:2004,Hu:2006,Holder:2006} and applied to data in, e.g.,  \citet{Baxter:2016, Chiu:2020}.

As a result of assembly bias, the linear bias parameter is also a function of halo concentration \citep{Wechsler:2006}.  Thus, the inference of the linear bias parameter for a sample of SZ-selected clusters will be impacted by the correlation between $Y$ and $c$.    We now estimate the size of the error that results from ignoring the $Y$-$c$ correlation.

Assuming that the linear bias is uniquely specified by $b(M,c)$, then for a $Y$-selected sample, the predicted mean linear bias is 
\begin{dmath}
    \bar{b}^Y = \int_{Y_{\rm min}}dY \int dM \int dc \, b(M, c)  P(Y | M,c) P(M,c) .
\end{dmath}
We wish to compare this prediction to one that ignores the concentration-dependence of $Y$.  To do this, we replace $P(Y|M,c)$ in the above expression by $P(Y|M) = \int dc \,P(Y|M,c)P(c|M)$, effectively marginalizing over $c$.  We refer to the resultant quantity as $\bar{b}^Y_{\rm noc}$.

We adopt the $b(M,c)$ relation from \citet{Wechsler:2006}:
\begin{equation}
b(M,c) = \langle b| M \rangle p(\tilde{m}) + q(\tilde{m})c' + 1.61(1-p(\tilde{m}))(c')^2
\end{equation}
with 
\begin{eqnarray}
p(\tilde{m}) &=& 0.95 + 0.042\ln (\tilde{m}^0.33) \\
q(\tilde{m}) &=& 0.1  - \frac{0.22(\tilde{m}^{0.33} + \ln (\tilde{m}^{0.33}))}{1 + \tilde{m}^{0.33}} \\
&&+ 0.042\ln (\tilde{m}^0.33) \nonumber \\
c' &=& \log_{10} \tilde{c}_{\rm vir} / \sigma(\log_{10} c_{\rm vir}) \\
\tilde{c}_{\rm vir} &=& c_{\rm vir}/\langle c_{\rm vir} | M_{\rm vir} \rangle \\
\tilde{m} &=& M_{\rm vir}/M_{\star} \\
\sigma(\log_{10} c_{\rm vir}) &=& 0.14,
\end{eqnarray}
where $\langle b| M \rangle$ is the average linear bias at fixed $M$ which we compute using the expressions from \citet{Tinker:2010}, and $M_{\star}$ is the typical collapsing mass at redshift $z$.  We convert between various mass definitions using \texttt{Colossus} \citep{colossus}.

\begin{figure}
    \centering
    \includegraphics[scale = 0.5]{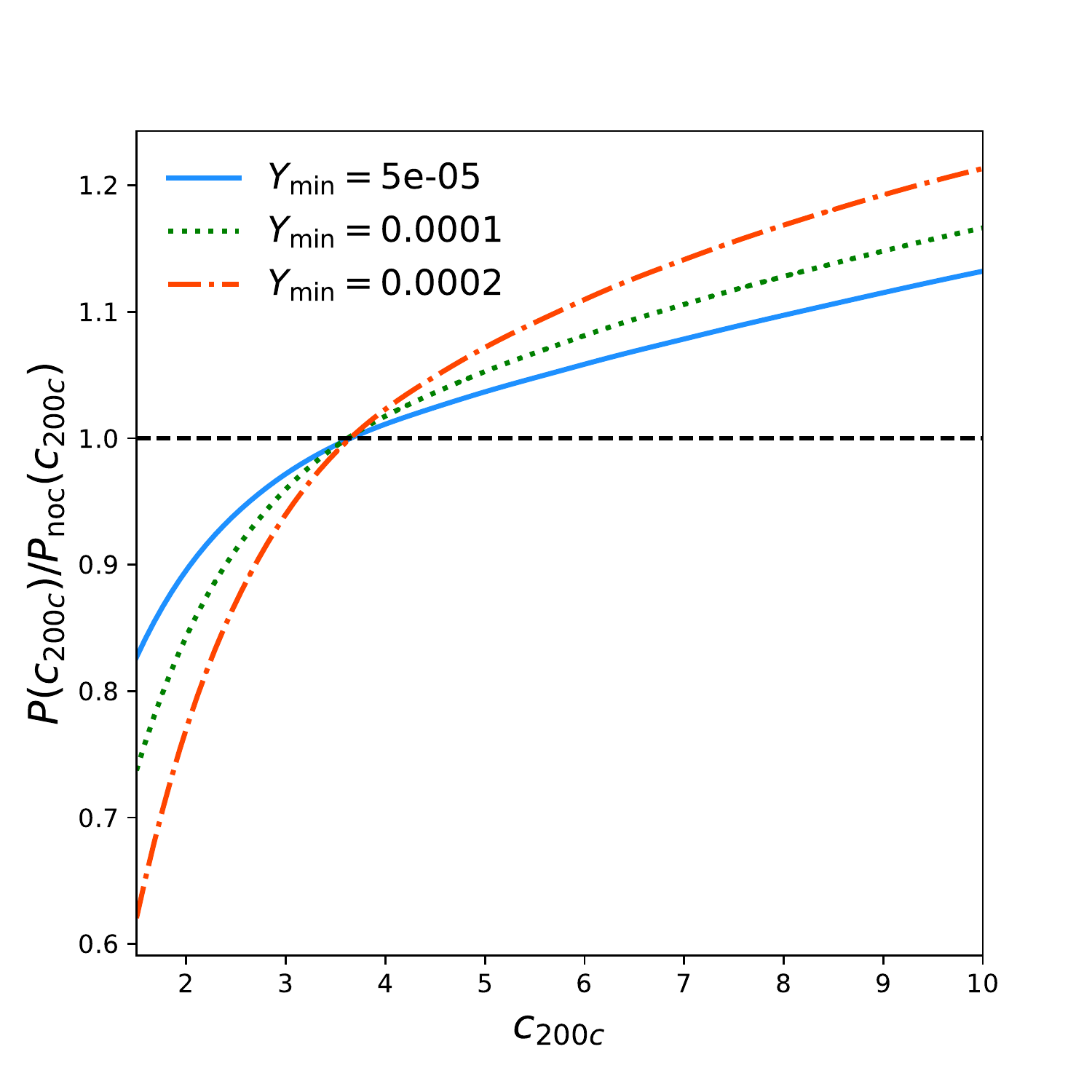}
    \caption{The change in the concentration distribution due to the concentration-dependence of $Y$, plotted for several choices of the minimum $Y$ used to define the cluster sample. The positive correlation between $Y$ and concentration means that a $Y$-selected sample becomes more likely to have halos with high concentration. }
    \label{fig:conc_distribution}
\end{figure}

\begin{figure}
    \centering
    \includegraphics[scale = 0.4]{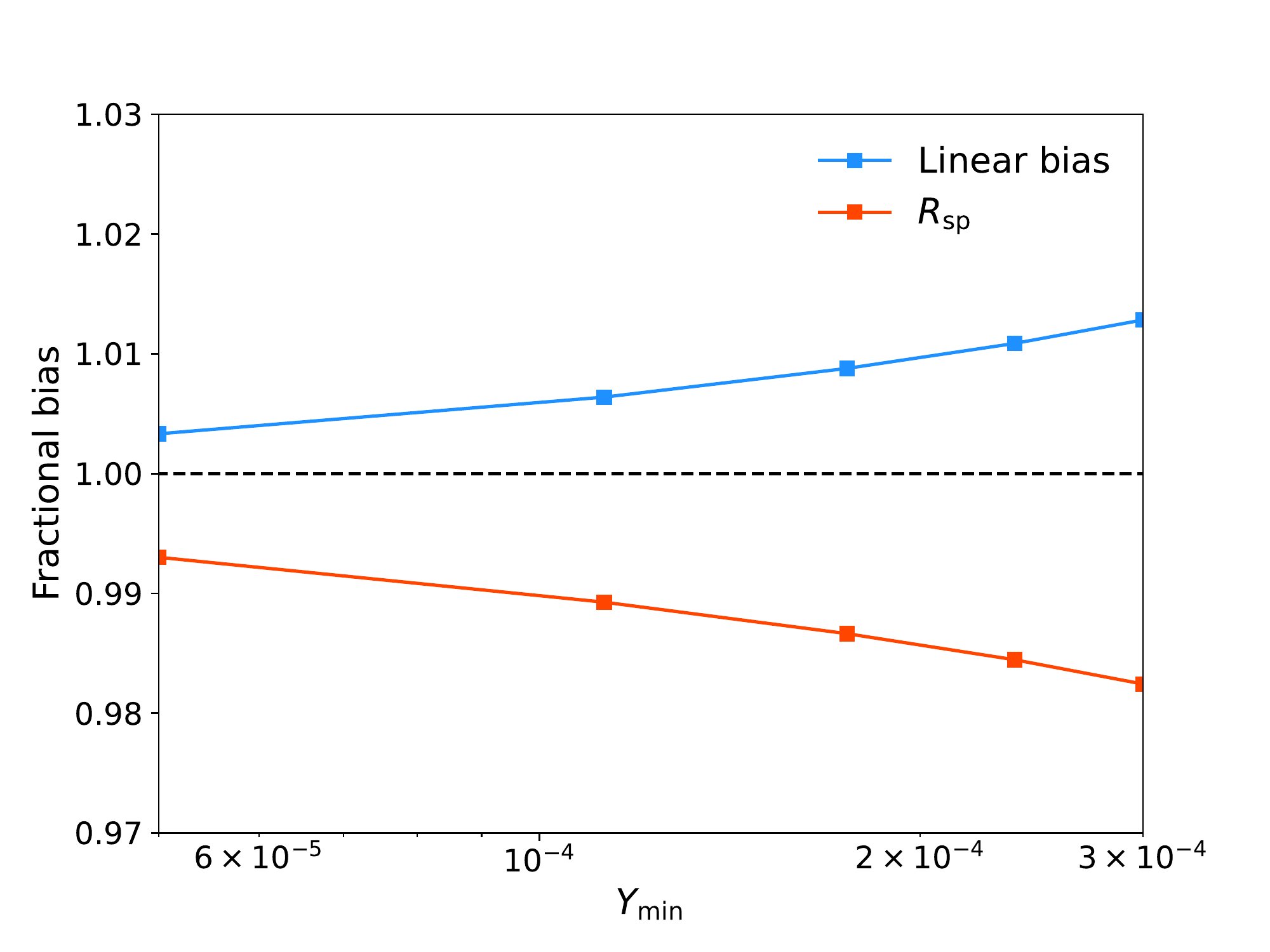}
    \caption{Fractional change in the expectation value of the linear bias parameter (blue) and splashback radius (red) as a result of ignoring the concentration dependence of $Y$ for an SZ-selected sample, plotted as a function of the minimum $Y$, $Y_{\rm min}$, used to define the sample.  A bias larger (smaller) than one means that the calculation of the quantity is over- (under-) estimated by ignoring the concentration dependence of $Y$.}
    \label{fig:biases}
\end{figure}

The impact on the modeled linear bias from ignoring the dependence of $Y$ on $c$ is shown in Fig.~\ref{fig:biases} (blue curve).  We plot the ratio of the calculation ignoring the concentration-dependence of $Y$ to the calculation that includes this dependence (i.e. $\bar{b}^Y_{\rm noc}/\bar{b}^Y$).   Ignoring the concentration-dependence of $Y$ shifts the predicted linear bias high relative to the correct calculation.  This is because the concentration distribution of a SZ-selected sample will be shifted to higher concentration values, as shown in Fig.~\ref{fig:conc_distribution}.  Higher concentration in turn corresponds to smaller linear bias, meaning that the incorrect calculation will overestimate the true linear bias.  We find that ignoring the concentration-dependence of $Y$ causes the linear bias parameter to be overestimated by at most 1\% for reasonable cluster selections.  This level of bias is likely to be smaller than measurement uncertainties with current and near-term cluster samples.

\subsubsection{Splashback radius}
\label{sec:splashback}

The splashback radius, $R_{\rm sp}$, varies with halo concentration.  As with the linear bias parameter, this means that ignoring the concentration dependence of $Y$ will result in a bias to the predicted $R_{\rm sp}$ for an SZ-selected sample.  Our calculation of this bias proceeds as in \S\ref{sec:clustering}, except that we replace $b(M,c)$ with $R_{\rm sp}(M,c)$.  Given that the measurements of $R_{\rm sp}$ from The300 agree with our truncation radius measurements, we calculate $R_{\rm sp}(M,c)/R_{200c}$ using Eq.~\ref{eq:R_t}.  We ignore scatter around the $R_{\rm sp}(M,c)$ relation, but have tested that reasonable levels of scatter have minimal impact on our results.  

The bias to the predicted $R_{\rm sp}$ of an SZ-selected sample from ignoring the concentration-dependence of $Y$ is shown with the red curve in Fig.~\ref{fig:biases}.  Since the splashback radius correlates positively with concentration, the bias is in the opposite direction as for the linear bias parameter: if one ignores the concentration dependence of $Y$, one will predict a value of $R_{\rm sp}$ for the sample that is too low.  We find that the bias in the splashback radius is at most 2\% for reasonable cluster selections.  The larger amplitude of this error relative to that for the linear bias parameter reflects the fact that the splashback radius depends more steeply on the concentration than the linear bias parameter.  Current uncertainties on the splashback radius measured by cross-correlating SZ-selected clusters with photometric galaxies are rough 20\% using samples of roughly 500 galaxy clusters \citep{Shin:2019}.  Since future CMB surveys like CMB Stage 4 \citep{CMBS4} are expected to find of order $10^5$ galaxy clusters, a 2\% bias could be significant for future splashback measurements with SZ-selected clusters.

\section{Summary}
\label{sec:discussion}

We have explored the relationship between halo concentration and SZ signal for massive galaxy clusters.  We find that polytropic gas models do an excellent job of describing how the inner SZ profiles ($R \lesssim 0.75 R_{200c}$) vary with concentration (Fig.~\ref{fig:profile_smallR}).  At larger projected distance ($R \gtrsim 0.75 R_{200c}$), however, these models do not perform as well (Fig.~\ref{fig:profile_largeR}).  As a result, the polytropic gas model from \citetalias{KomatsuSeljak2001} makes incorrect predictions for the relationship between halo concentration and integrated SZ signal at low concentration (see in particular the blue curve in the left panel of Fig.~\ref{fig:Y_vs_concentration}).  

In order to improve predictions for the $Y_{\Delta}$-$c_{\Delta}$ relation and better match the outer SZ profiles, we find it necessary to introduce a truncation of the halo pressure profiles.  Such truncation is motivated by self-similar infall models \citep{Bertschinger:1985} for which the radius of truncation is expected to coincide with the splashback radius \citep{Shi:2016}.  Indeed, we find that the  truncation radius must closely track the splashback radius in order to explain the $Y$-$c_{\Delta}$ relation.  Our model for the truncated pressure profile is provided in Eq.~\ref{eq:final_model}.  Relative to standard polytropic models that do not include such truncation, the model we have introduced provides an improved match to both the halo pressure profiles and the $Y$-concentration relation measured in simulations. Our simple model for the truncation of the pressure profile (i.e. that it is maximally steep), is not expected to provide a perfect description of the halo pressure profiles near the truncation radius.  However, even with this simple prescription, we find obtain significantly improved predictions of e.g. the  $Y$-$c_{\Delta}$ relation  (Fig.~\ref{fig:Y_vs_concentration}, red curves).

We use this model to assess whether ignoring the $Y_{\Delta}$-$c_{\Delta}$ relation can lead to biases in the inference of the linear bias parameter and splashback radii of SZ-selected cluster samples.  We find that such biases are small for the linear bias parameter --- generally less than 1\% --- making them largely irrelevant for analyses with current and near-term samples of SZ-selected clusters.  The bias to the inference of the splashback radius is somewhat larger, and may be important for near-term measurements (Fig.~\ref{fig:biases}).

An important caveat of this work is that we have restricted our attention to a fairly narrow range of halo masses ($M \gtrsim 5 \times 10^{14}\,\mathrm{M_{\odot}}$ and redshift $z = 0.193$).  This restriction is unlikely to qualitatively impact our main results, but it is possible/likely that outside of this range, there may be some variation in the parameters of our simple model.  We postpone a global fit of our model to halos across a broader range of redshifts and masses to future work.

\section{Acknowledgements}

We thank Chihway Chang, Bhuvnesh Jain, Andrey Kravtsov and Heidi Wu for fruitful discussions related to this work.  We also thank an anonymous referee for their helpful suggestions.  

TS is supported by the U.S. Department of Energy grant DE-SC0023387.

WC is supported by the STFC AGP Grant ST/V000594/1 and the Atracci\'{o}n de Talento Contract no. 2020-T1/TIC-19882 granted by the Comunidad de Madrid in Spain. He further thanks the Ministerio de Ciencia e Innovación (Spain) for financial support under Project grant PID2021-122603NB-C21 and the European Research Council’s: HORIZON-TMA-MSCA-SE grant number 101086388, i.e. the LACEGAL-III (Latin American Chinese European Galaxy Formation Network) project.

\section{Data availability}

This manuscript was developed using data from \thethreehundred\ galaxy clusters sample. The data is available on request following the guideline of \thethreehundred\ collaboration (\url{https://www.the300-project.org}). The data used to make the figures shown in this work are available upon request.

\bibliographystyle{mnras}
\bibliography{thebib}

\appendix

\bsp	
\label{lastpage}
\end{document}